\documentclass[%
nofootinbib,
 amsmath,amssymb,
 aps,
]{revtex4-2}

\usepackage{graphicx}
\usepackage{dcolumn}
\usepackage{bm}
\usepackage{hyperref}

\begin{document}

\title{MATOQ: a Monte Carlo Simulation of Electron Transport in Environmental-friendly Gas Mixtures for Resistive Plate Chambers} 

\author{Antonio Bianchi} \email{Corresponding author: antonio.bianchi@cern.ch}
%
\affiliation{CERN}

\begin{abstract}

The increasing interest in environmentally friendly gas mixtures for gaseous particle detectors, especially tetrafluoropropene-based gas mixtures for Resistive Plate Chambers (RPCs), has prompted the need for simulating electron transport coefficients and reaction rates in these mixtures in recent years. MATOQ is a Monte Carlo simulation program that calculates electron transport parameters, specifically designed for studying and optimizing environmental-friendly gas mixtures for RPCs. Unlike other existing codes, MATOQ allows for the simulation of electron avalanches by including the effect of space charge electric field, which can significantly impact the avalanche evolution in gaseous detectors such as RPCs.

After the validation of the MATOQ simulation in the temporal and spatial growth configurations, we present the electron transport coefficients and the reaction rates in tetrafluoropropene-based gas mixtures, which may represent a valid alternative to the standard gas mixtures currently used for RPCs.

\end{abstract}
\maketitle









\section{Introduction}

Resistive Plate Chambers (RPCs) are gaseous particle detectors used in high-energy physics experiments \cite{santonico1981development, bruno2004resistive, abbrescia2012eee, zeballos1996new} and medical imaging applications \cite{amaldi2015development, crespo2013resistive}. They consist of two parallel plates made of high-resistivity materials with a gap between them filled with a gas mixture. Tetrafluoroethane (C$_{2}$H$_{2}$F$_{4}$) is generally the main component of the gas mixtures for RPCs. This gas is typically mixed with quench gases such as isobutane (\textit{i}-C$_{4}$H$_{10}$) and sulfur hexafluoride (SF$_{6}$) in various proportions to optimize the performance of RPCs for specific applications.

In view of supporting the transition to a green economy and fighting climate change, recent regulations of the European Union have prohibited the use of C$_{2}$H$_{2}$F$_{4}$ for many applications, since it is a greenhouse gas. Indeed, the global warming potential of C$_{2}$H$_{2}$F$_{4}$ is about 1430 \cite{ipcc}. This means that the impact of this gas on the greenhouse effect is estimated to be 1430 times higher than an equivalent mass of carbon dioxide (CO$_{2}$) in the atmosphere. Although there are no European regulations restricting the use of C$_{2}$H$_{2}$F$_{4}$ for scientific applications, some research teams \cite{abbrescia2016eco, guida2016characterization, liberti2016further, bianchi2019characterization} have explored the possibility of replacing C$_{2}$H$_{2}$F$_{4}$ with more environmentally friendly gases. Many experimental studies are currently focused on measuring the performance of RPCs by replacing the current C$_{2}$H$_{2}$F$_{4}$-based gas mixture with environmental-friendly alternatives \cite{bianchi2019characterization, proto2022new, abbrescia2016preliminary, rigoletti2020studies}. Recently, some encouraging results have been obtained by replacing the C$_{2}$H$_{2}$F$_{4}$ with tetrafluoropropene. Tetrafluoropropene (C$_{3}$H$_{2}$F$_{4}$) has a chemical composition similar to C$_{2}$H$_{2}$F$_{4}$, but it seems much more electronegative than C$_{2}$H$_{2}$F$_{4}$ \cite{chachereau2016electron, bianchi2020studies}. This would result in too high operating voltages for RPCs that are not compatible with the already existing power supply systems. Instead of replacing C$_{2}$H$_{2}$F$_{4}$ with only C$_{3}$H$_{2}$F$_{4}$, binary mixtures of C$_{3}$H$_{2}$F$_{4}$ and CO$_{2}$ may be a feasible alternative \cite{bianchi2019characterization, proto2022new, abbrescia2016preliminary, rigoletti2020studies}. However, purely experimental studies require a large number of trials to identify gas mixtures that yield satisfactory performance of RPCs.

The simulation of electron transport coefficients and reaction rates under the influence of the electric field can assist in the selection of the most promising eco-friendly gas mixtures for RPCs. In recent years, some Monte Carlo programs to simulate electron transport in gases under the influence of a static and uniform electric field have been developed for specific applications. One of the most widely used codes is MAGBOLTZ that was developed by S. Biagi in the 1990s and has been regularly updated \cite{biagi1999monte}. This open-source code, written in FORTRAN, is still used in the field of gaseous particle detectors. However, one of the most significant limitations is that the input electron collision cross sections for all gases are deeply embedded in the MAGBOLTZ code. This makes it challenging to implement new sets of electron collision cross sections, like those for C$_{3}$H$_{2}$F$_{4}$ \cite{bianchi2021electron}, and modifying the code for specific purposes can be complicated. Some attempts have been recently done to implement more user-friendly Monte Carlo simulations similar to MAGBOLTZ. In particular, the METHES program \cite{rabie2016methes} overcomes the limitation of MAGBOLTZ in simulating electron transport in gases not yet included in the internal database. Indeed, different sets of electron collision cross sections can be easily adopted as input in METHES. However, the execution of METHES requires a commercial license of MATLAB \cite{rabie2016methes}. In addition, it is important to note that MAGBOLTZ and METHES do not account for the effect of the space charge electric field, which can significantly impact the avalanche evolution, especially in RPCs with narrow gas gaps.

This paper describes the MATOQ program that is a Monte Carlo simulation focused on the calculation of electron transport coefficients and reaction rates in any gas mixture of interest under the influence of the electric field. This is obtained by simulating the temporal and spatial growth of electron avalanches along gas gaps. In addition, MATOQ allows simulating the electron avalanche growth under the influence of a static applied electric field together with the space charge electric field that changes depending on the avalanche evolution along a given gas gap. This aspect sets the MATOQ program apart from all other available Monte Carlo simulations.

MATOQ is implemented in the programming language C++, which facilitates its usage and customization in various research fields where C++ is commonly used, such as in the simulation of gaseous particle detectors. MATOQ is compatible with the file format of electron collision cross sections adopted by the open-access Plasma Data Exchange Project (LXCat) \cite{lxcat_project} in order to allow the user to easily control all input parameters. Moreover, MATOQ is interfaced with OpenMP \cite{openmp} to enable multi-thread execution, and with the ROOT program \cite{root_documentation} for the graphical representation of simulation results.

The paper is organized as follows. In section \ref{sec:space_charge} we describe how the space charge electric field may play a significant role in the electron avalanche growth. The methodology to simulate the electron transport in gases is presented in section \ref{sec:methods}, while the simulation of collisions between electrons and neutral gas molecules is examined in section \ref{sec:types}. Sections \ref{sec:temporal} and \ref{sec:spatial} describe the temporal and spatial growth configurations of MATOQ, whereas the simulation of the avalanche growth under the influence of the applied electric field together with the space charge electric field is detailed in section \ref{sec:spatial_space_charge}. In section \ref{sec:electron_transport}, we compare the electron transport coefficients and reaction rates in pure C$_{2}$H$_{2}$F$_{4}$, pure C$_{3}$H$_{2}$F$_{4}$, and in binary mixtures of C$_{3}$H$_{2}$F$_{4}$ and CO$_{2}$. Furthermore, we describe how the avalanche size in a narrow-gap RPC is affected when C$_{2}$H$_{2}$F$_{4}$ is substituted by C$_{3}$H$_{2}$F$_{4}$. Finally, conclusions are drawn in section \ref{sec:conclusion}.

\section{Space charge electric field}\label{sec:space_charge}
Free charged particles gain energy in gases under the influence of an electric field. Since the ion mobility is generally three orders of magnitude lower than that of electrons \cite{abbrescia_book}, the velocity of ions in gases is generally negligible in comparison to that acquired by electrons. As a result, the number of electrons grows exponentially by giving rise to an electron avalanche. Indeed, under the influence of the electric field, electrons can gain enough energy to ionize a certain number of gas molecules along their drift towards the anode. On the contrary, ions generally drift towards the cathode without playing a fundamental role in the charge multiplication due to their low energy. In this work, the motion of ions is not simulated since they move much slower than electrons. Nevertheless, the motion of ions can be easily implemented in MATOQ by including the ion mobility of each species in the gas mixture of interest.  

During the avalanche growth in gases, electrons and ions are partially overlapped in space while they move towards the opposite electrodes. This generates an electric field, generally called space charge electric field, that is superimposed on the applied electric field. As a consequence, the applied electric field turns out to be reduced in the middle of the electron avalanche because it is lowered by the space charge electric field between electrons and ions. On the contrary, the applied electric field is strengthened in the upstream and downstream avalanche as it is reinforced by the space charge electric field. This causes a non-uniform electric field during the avalanche evolution in gas gap that depends on free charges and their positions in time. One of the common devices where space charge effects can have a significant impact is gaseous particle detectors, in particular narrow-gap RPCs. \cite{abbrescia_book, mrpc}. 

MATOQ allows for calculating or excluding space charge effects in simulations depending on the desired outcome. In sections \ref{sec:temporal} and \ref{sec:spatial}, the space charge electric field is not assessed in order to validate the MATOQ calculations with the results of MAGBOLTZ, in which the space charge effects cannot be simulated. On the contrary, in section \ref{sec:spatial_space_charge}, the space charge electric field is considered for simulating the average avalanche size as a function of the applied electric field in an RPC with a gas gap of 0.1 mm. This is done to enable a comparison between the MATOQ results and those obtained by the Lippmann et al.'s model, which can be used to calculate the electron avalanche size in RPCs \cite{lippmann2004space}.

\section{Methods}\label{sec:methods}
The MATOQ simulation program tracks the motion of electrons for the entire duration of the simulation, while ions are considered motionless during electron avalanche development due to their much lower mobility compared to electrons. The position $\vec{r}$ and the velocity $\vec{v}$ of a free electron in a gas mixture under the influence of the applied electric field $\vec{E}$ is determined according to the following equations:
\begin{equation}
    \vec{r} \rightarrow \vec{r} + \vec{v} \Delta t + \frac{1}{2} \frac{e\vec{E}}{m_{e}} \Delta t^{2} \,\, \textup{and} \,\,
    \vec{v} \rightarrow \vec{v} + \frac{e\vec{E}}{m_{e}} \Delta t
\end{equation}
where $\Delta t$ is the time step of the simulation, while $m_{e}$ and $e$ are the electron mass and charge, respectively. The kinetic energy $\varepsilon$ of an electron with velocity $\vec{v}$ is given by:
\begin{equation}
    \varepsilon = \frac{1}{2} m_{e} |\vec{v}|^{2}
\end{equation}

The choice of an appropriate time step $\Delta t$ to perform the MATOQ simulation is determined by the null-collision technique \cite{null1, null2}. According to this technique, the probability $P(\Delta t)$ of time steps higher than $\Delta t$ is:
\begin{equation}\label{eq:formula_probability_cap5}
    P(\Delta t) = e^{-\int_{0}^{\Delta t} \nu(|\vec{v}(t)|) \, d t }
\end{equation}
where $t$ is the time and $\nu$ is the collision frequency, which depends on the electron velocity $\vec{v}$. Indeed, the collision frequency $\nu$ can be expressed as:
\begin{equation}
    \nu(|\vec{v}|) = N \sigma(|\vec{v}|) |\vec{v}|
\end{equation}
where $N$ denotes the number of gas molecules per unit volume, which is assumed constant in space and time in all MATOQ simulations, and $\sigma$ is the cross section of each individual process that can take place in the gas mixture. For a gas mixture consisting of $M$ components with respective concentrations $c_{m}$, the total cross section $\sigma_{tot}$ is given by:
\begin{equation}
    \sigma_{tot}(|\vec{v}|) = \sum_{m}^{M} \sum_{i}^{I} c_{m} \sigma_{m,i}(|\vec{v}|)
\end{equation}
where $m$ is the index of each gas component, while the index $i$, ranging from 1 to $I$, corresponds to each individual electron collision process that can occur in the gas component $m$. Using the null-collision technique, a constant trial collision frequency ($\nu^\prime$) is introduced and assumed higher than the total collision frequency $\nu_{tot}$ in the whole energy range of interest. As a consequence, the expression of $\nu^\prime$ is:
\begin{equation}\label{eq:nuprimoenu}
    \nu^\prime > \textup{max}(\nu_{tot}(|\vec{v}|)) = \textup{max}(N  \sigma_{tot}(|\vec{v}|) |\vec{v}|)
\end{equation}
The total cross section $\sigma_{tot}$ is evaluated in MATOQ between 0 eV and 100 eV, whereas the trial collision frequency $\nu^\prime$ is assumed three times higher than $\nu_{tot}$. As a result, the introduction of a constant trial collision frequency $\nu^\prime$ that is independent of the electron velocity $\vec{v}$ gives the possibility to recast the probability $P(\Delta t)$ as follows:
\begin{equation}
    P(\Delta t) = e^{-\nu^\prime \Delta t}
\end{equation}
Therefore, the selection of the time step can be determined by the generation of a random number. Using the inverse transformation method for the distribution $P(\Delta t)$, the time step $\Delta t$ is calculated as follows:
\begin{equation}\label{eq:formula_random_number}
    \Delta t = - \frac{1}{\nu^\prime} \ln(s)
\end{equation}
where $s$ is a random number generated from a uniform distribution in the range (0, 1].

For each electron, identified by the index $k$, the possibility that a real collision may occur is checked after every time step $\Delta t$. The number $L$ of all possible processes at the corresponding velocity $\vec{v}_{k}$ is determined to define a vector with $L$+1 items for each electron \cite{rabie2016methes}. Subsequently, the single item $C_{l}$ of the vector is initialized as:
\begin{equation}\label{eq:formulaIndici}
    C_{l} = C_{l-1} + \frac{N \cdot c_{m} \cdot \sigma_{m, l}(|\vec{v}_{k}|) \cdot |\vec{v}_{k}|}{\nu^\prime}
\end{equation}
where $C_{l}$ is the occurrence probability of $l$-th electron collision process. According to equation \ref{eq:formulaIndici}, the content of each item is cumulatively summed to obtain an integral function that monotonically increases. Since a null-collision can occur, the total number of items in the vector is $L$+1 and each collision frequency $N \cdot c_{m} \cdot \sigma_{m, l}(|\vec{v}_{k}|) \cdot |\vec{v}_{k}| $ is normalized by $\nu^\prime$, which is higher than the total collision frequency $\nu_{tot}$, in accordance with the equation \ref{eq:nuprimoenu}. By means of this normalization, the last item of the vector represents the occurrence probability of a null-collision, in which no real collision occurs.

\section{Types of electron collisions}\label{sec:types}
For each type of electron collision, MATOQ needs the corresponding cross section as a function of the incident electron energy as input. Since there are different sets of electron collision cross sections present in the literature for the most common gases, MATOQ is compatible with the data format used in the LXCat database, which includes a vast collection of sets for various gases \cite{lxcat_project}. This is the same approach adopted in METHES, whereas MAGBOLTZ utilizes a database included in the code and updated and reviewed periodically by its developer.

In MATOQ, simulations of electron-neutral gas molecule collisions are carried out, while interactions between electrons and between electrons and ions are not considered. The possible electron collisions that can be simulated include: (a) elastic processes, where the interaction involves an electron and a neutral gas molecule in both the initial and final states; (b) excitations, where an incident electron transfers part of its energy to a neutral gas molecule, promoting it to an excited state. However, secondary effects such as photon emission are not considered in MATOQ; (c) ionization and attachment events, where an incident electron ionizes a neutral gas molecule or is captured by it, respectively.

In the LXCat database, cross sections for every type of electron collision are presented as tables of values that vary as a function of the incident electron energy. In MATOQ, each cross section value is linearly interpolated with the subsequent value in the table to obtain a continuous function that spans the entire energy range of interest.

In order to select the type of electron collision to simulate in MATOQ, a random number $r$ is generated from a uniform distribution ranging from 0 to 1 for each time step $\Delta t$. If $r$ is smaller than the content of $C_{1}$ in equation \ref{eq:formulaIndici}, the first collision process is simulated. If $r$ is larger than the ($l$-1)-th item and smaller than the $l$-th item, the $l$-th process is simulated. If $r$ is higher than the content of $C_{L}$, a null-collision is selected, which means that the electron continues its free motion without any interactions with the gas medium \cite{rabie2016methes}.

In the case of elastic collisions, isotropic scattering is assumed in the simulation, meaning that the polar angle $\theta$ and the azimuthal angle $\phi$ are determined as follows \cite{rabie2016methes}:
\begin{equation} \label{eq:angoli_MATOQ}
    \theta = \arccos(1 - 2 r_{1}) \,\, \textup{and} \,\,
    \phi = 2\pi r_{2}
\end{equation}
where $r_{1}$ and $r_{2}$ are random numbers generated from a uniform distribution ranging from 0 to 1. In addition to the simulation of the resulting electron trajectory, it is necessary to evaluate the energy loss of the incident electron for each elastic collision. Since gas molecules are assumed to rest in the laboratory system for simplicity, the energy loss $\Delta\epsilon$ of the incident electron after the elastic scattering with the neutral gas molecule of mass $M$ is given by \cite{rabie2016methes, landau}:
\begin{equation}\label{eq:bilancioenergetico}
    \Delta \epsilon = \frac{1}{2} \epsilon \frac{m_{e}}{M} (1 - \cos(\theta))
\end{equation}
where $\epsilon$ is the incident electron energy, $m_{e}$ is the electron mass and $\cos(\theta)$ is given by the equation \ref{eq:angoli_MATOQ}. On the contrary, the electron energy loss $(\Delta\epsilon)^{*}$ after an excitation or ionization process is given by:
\begin{equation}\label{eq:bilancioenergeticoion}
    (\Delta\epsilon)^{*} = \epsilon^{*}
\end{equation}
where $\epsilon^{*}$ is the energy threshold of the specific event to simulate. In other terms, the incident electron energy is assumed to be reduced by the minimum energy required to excite or ionize the gas molecule, depending on the type of process simulated. The scattering of the incident electron in excitation and ionization processes is considered isotropic as in the case of elastic processes. The values of $\epsilon^{*}$ used in MATOQ are obtained from the LXCat database along with the electron collision cross sections. More details on simulating ionization and attachment processes are given in sections \ref{sec:temporal} and \ref{sec:spatial}, depending on the chosen temporal or spatial growth configuration. In the case of a null-collision, the electron does not interact with the gas medium and continues its motion.

\section{Temporal growth configuration}\label{sec:temporal}
In the temporal growth configuration of MATOQ, trajectories and collisions of an ensemble of electrons are simulated in an infinite gas volume. The number of electrons remains constant for the entire duration of the simulation. This configuration enables the determination of various electron properties such as their mean energy, drift velocity, and ionization and attachment coefficients.

MATOQ implements the same technique used in METHES \cite{rabie2016methes} to maintain a constant number of electrons at every time step $\Delta t$. In the case of an ionization event, an additional electron is added to the electron ensemble, while a different electron is randomly removed. The new electron is simulated from the initial position where the ionization occurred.  The remaining electron energy, calculated as the difference between the energy of the incident electron and the ionization energy of the involved gas molecule (according to equation \ref{eq:bilancioenergeticoion}), is equally shared between the two electrons resulting from the ionization collision. On the contrary, in the case of attachment processes, the attached electron is removed from the electron ensemble, and an additional electron is added to the same position where the attachment event occurred. The direction and energy of the new electron are assumed to be the same as those of an electron randomly selected from the ensemble.

\begin{figure}[ht]
    \centering
    \includegraphics[width=0.50\textwidth]{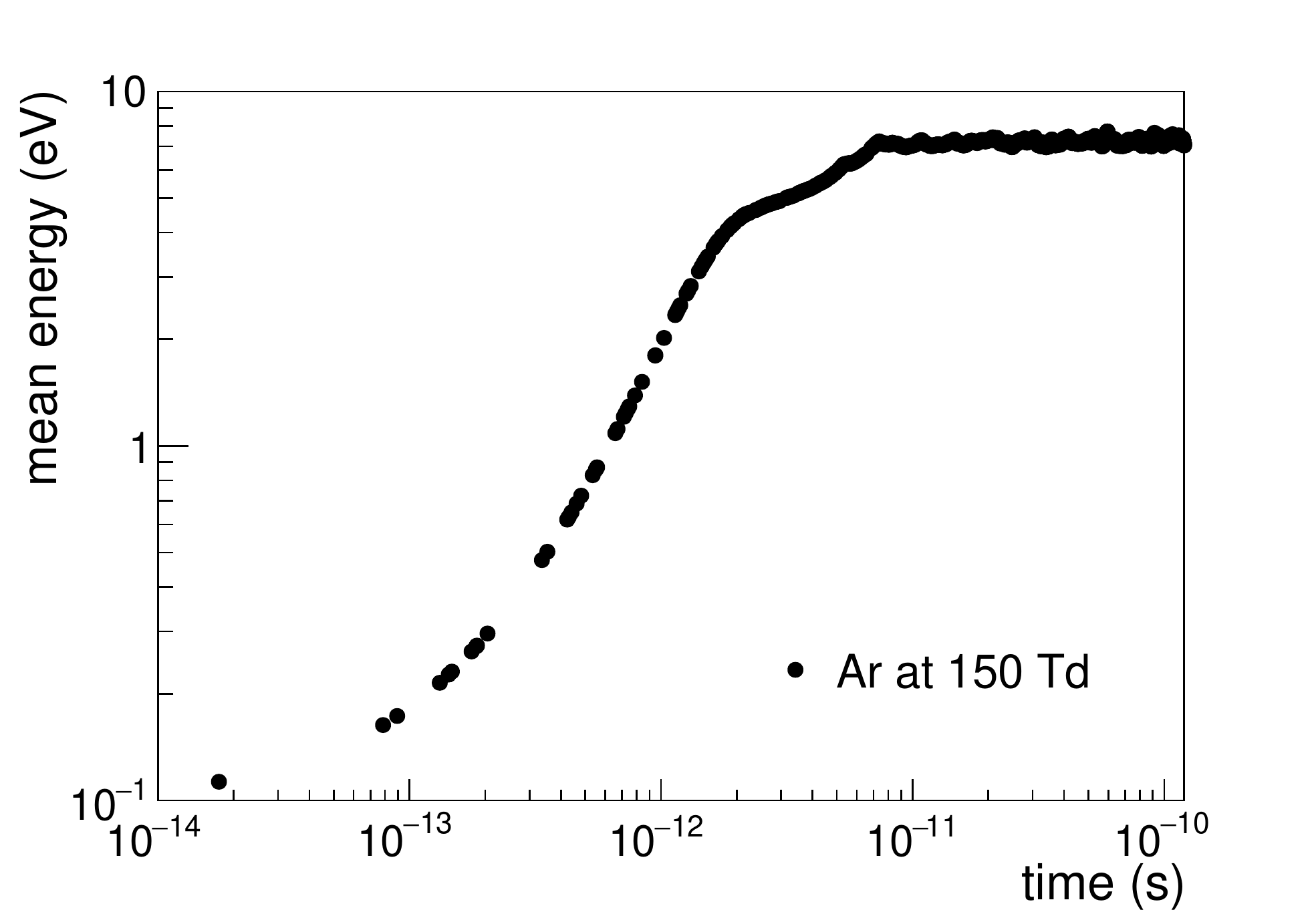}
    \caption{Mean energy of 10$^{5}$ electrons as a function of the time in pure Ar at 150 Td.}
    \label{fig:plot1}
\end{figure}

The reliable calculation of electron transport coefficients and reaction rates is only possible after the electron ensemble has reached a steady state regime, where the mean electron energy remains constant, except for statistical fluctuations. All electrons in the MATOQ simulation start with an initial energy of 0.1 eV. Figure \ref{fig:plot1} shows the time evolution of the mean energy of 10$^{5}$ electrons in pure argon (Ar) at a reduced electric field of 150 Td\footnote{1 Td = 10$^{-21}$ V$\cdot$m$^{2}$}. After approximately 10$^{-11}$ s, the mean electron energy reaches a constant value of around 7 eV, with small statistical variations ($<\sim$0.5 eV).

The instantaneous mean electron energy $<\!\varepsilon(t)\!>$ at the time $t$ is given by:
\begin{equation}\label{eq:energia_istantanea_MATOQ}
    <\!\varepsilon(t)\!> = \frac{1}{2} m_{e} \frac{1}{K} \sum_{k = 1}^{K}|\vec{v}_{k}(t)|^{2}
\end{equation}
where $k$ indicates the $k$-th electron with velocity $\vec{v}_{k}(t)$, whereas $K$ is the total number of electrons. Similarly to the calculation of the instantaneous mean electron energy, the instantaneous mean electron velocity $\vec{v}(t)$ at the time $t$ is given by:
\begin{equation}\label{eq:cap5_formula3}
\vec{v}(t) = \frac{1}{K}\sum_{k = 1}^{K}\vec{v}_{k}(t)
\end{equation}

Since the number of electrons does not change in the temporal growth configuration of MATOQ, the number of ionizations and attachments increases linearly in time \cite{rabie2016methes}. Therefore, the ionization ($\nu_{ion}(t)$) and electron attachment ($\nu_{att}(t)$) coefficients at the time $t$ are given by:
\begin{equation}\label{eq:cap5_formula5}
    \nu_{ion}(t) = \frac{N_{ion}(t) - N_{ion}(t_{0})}{K |\vec{v}(t)|  (t - t_{0})} \,\, \textup{and} \,\,
    \nu_{att}(t) = \frac{N_{att}(t) - N_{att}(t_{0})}{K |\vec{v}(t)|  (t - t_{0})}
\end{equation}
where $N_{ion}(t_{0})$ and $N_{att}(t_{0})$ are the number of the ionizations and electron attachments at the time instant $t_{0}$, respectively, whereas $N_{ion}(t)$ and $N_{att}(t)$ correspond to the number of ionization and attachment events at time instant $t$ with $t > t_{0}$.

To accurately evaluate all electron transport coefficients and reaction rates, the electron ensemble must reach a steady state. Therefore, $<\!\varepsilon(t)\!>$, $\vec{v}(t)$, $\nu_{ion}(t)$ and $\nu_{att}(t)$ are only calculated after this condition has been met. The mean electron energy $<\!\varepsilon\!>$ and the mean electron velocity $\vec{v}$ are determined in MATOQ by averaging all respective values of $<\!\varepsilon(t)\!>$ and $\vec{v}(t)$, sampled at each time step $\Delta t$ after reaching the steady state. In this work, the drift velocity $v_{drift}$ is defined as the component of velocity $\vec{v}$ along the direction of the applied electric field $\vec{E}$. The ionization and attachment rates, $\nu_{ion}$ and $\nu_{att}$, are calculated by counting the respective processes every 1000 time steps after reaching the steady state. This ensures that enough ionization and attachment events occur between samplings. The values of $<\!\varepsilon\!>$, $\vec{v}$, $\nu_{ion}$ and $\nu_{att}$ are subject to statistical fluctuations, and their uncertainties are estimated by computing the standard deviation of the corresponding values obtained at each sampling.

The temporal growth configuration in MATOQ simulation ends after a specified number of real collisions, which is selected by the user. Increasing the number of real collisions improves the accuracy of results in Monte Carlo simulations, but it also increases the computation time \cite{biagi1999monte, rabie2016methes}. In MATOQ, tens of millions of real collisions generally result in a reasonable computation time with satisfactory accuracy.

\begin{figure}[ht]
    \centering
    \includegraphics[width=0.95\textwidth]{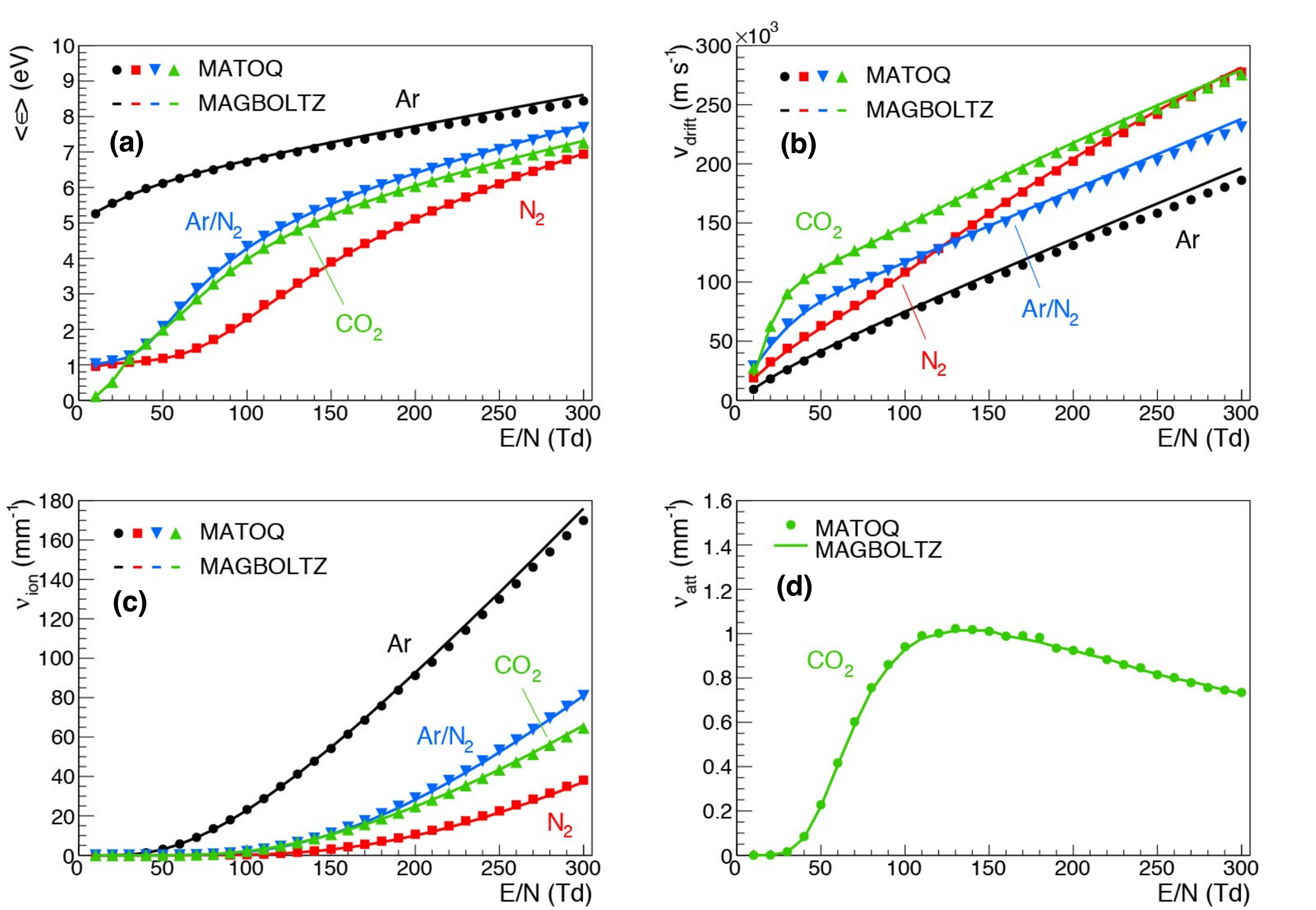}
    \caption{Values of average electron energy $<\!\varepsilon\!>$ (a), drift velocity $v_{drift}$ (b), ionization coefficient $\nu_{ion}$ (c) and attachment coefficient $\nu_{att}$ (d) as a function of the reduced electric field $E/N$ in pure Ar, N$_{2}$, CO$_{2}$ and in the gas mixture of 50\% Ar and 50\% N$_{2}$. Some statistical error bars are hidden by markers.}
    \label{fig:plot2}
\end{figure}

To validate the MATOQ simulation, we compare the calculated values of $<\!\varepsilon\!>$, $v_{drift}$, $\nu_{ion}$, and $\nu_{att}$ as a function of the reduced electric field $E/N$ with MAGBOLTZ results in both pure Ar and nitrogen (N$_{2}$) gases. Additionally, we examine the accuracy of MATOQ by comparing the results with MAGBOLTZ in a binary mixture of 50\% Ar and 50\% N$_{2}$ as well as in pure CO$_{2}$ where electron attachments can occur, unlike in Ar and N$_{2}$. The electric field is assumed uniform in the gas medium in all these cases. No space charge effects are considered here. Figure \ref{fig:plot2} shows values of $<\!\varepsilon\!>$, $v_{drift}$, $\nu_{ion}$ and $\nu_{att}$ as a function of $E/N$ in pure Ar, N$_{2}$, CO$_{2}$ and in the gas mixture composed of 50\% Ar and 50\% N$_{2}$. For each value of $E/N$, MATOQ simulation results are obtained with 10$^{5}$ electrons and the maximum number of real collisions is set equal to 4$\cdot$10$^{8}$ as in MAGBOLTZ. The steady state is assumed to be reached in MATOQ after 1$\cdot$10$^{8}$ real collisions. Sets of electron collision cross sections used for simulations are specified in the appendix.

The comparison between MATOQ and MAGBOLTZ calculations in all four gas mixtures shows a good agreement, as displayed in figure \ref{fig:plot2}. The relative error\footnote{The relative error $r$ of the generic parameter $A$ is defined in percentage as follows:
\begin{equation}
    r_{A(E/N)} = \left(\frac{A_{MQ}(E/N) - A_{MZ}(E/N)}{A_{MZ}(E/N)}\right) \cdot 100
\end{equation}
where A$_{MQ}$($E/N$) and A$_{MZ}$($E/N$) are the values of the parameter $A$($E/N$) calculated at the value $E/N$ with MATOQ and MAGBOLTZ, respectively.} in the values of $<\!\varepsilon\!>$, $v_{drift}$, $\nu_{ion}$ and $\nu_{att}$ is below 1\% for all gas mixtures tested, except for pure Ar at $E/N$ values above 200 Td, where the relative error increases up to about 5\%. This discrepancy is likely due to the assumption in MATOQ that the energy resulting from ionization collisions is equally shared between the two electrons in the final state, which may lead to less accurate results compared to MAGBOLTZ. This effect is more pronounced in Ar at higher $E/N$ values where ionization events are more frequent compared to the other gases.

\section{Spatial growth configuration}\label{sec:spatial}
The spatial growth configuration simulates an electron avalanche in an infinite gas volume, with a certain number of initial electrons and a given $E/N$ value. The number of electrons is not fixed during the simulation. An additional electron is added to the avalanche upon ionization, while the trajectory of an electron is not simulated anymore when it becomes captured by a gas molecule. This configuration allows us to determine the ionization ($\alpha$) and attachment ($\eta$) Townsend coefficients. In the presence of ionization and attachment processes, the number of electrons $n(x)$ at the distance $x$ is given by:
\begin{equation}\label{eq:townsend}
    n(x) = n_{0} e^{(\alpha - \eta) x} = n_{0} e^{\alpha_{eff} x}
\end{equation}
where $n_{0}$ is the initial number of electrons while $\alpha$ and $\eta$ are the ionization and attachment Townsend coefficients, respectively. The difference between $\alpha$ and $\eta$ is usually named effective ionization Townsend coefficient $\alpha_{eff}$.

The spatial growth configuration of MATOQ simulates the electron transport and energy transfer after collisions in the same way as the temporal growth configuration, with the exception of ionization and attachment processes. Upon ionization, an additional electron with a random initial direction is added to the simulation, and the two electrons resulting from the collision share the remaining energy equally. The trajectory of the new electron starts from the same position as the incident electron. In the case an electron becomes attached to a gas molecule, its motion is not simulated any further. It should be noted that the spatial growth configuration may not be effective in the presence of high attachment coefficients, as all electrons in the avalanche could become attached before the simulation concludes.

The evaluation of $\alpha$ and $\eta$ for a given $E/N$ value in MATOQ is performed by counting the number of electrons that cross a series of virtual planes, placed at the same distance apart and perpendicular to the electric field direction. The position of the virtual planes is determined based on the positions of the slower electron at the beginning of the steady state and at the end of the simulation. To obtain reliable values of $\alpha$ and $\eta$, tens of virtual planes are usually sufficient. During the simulation, the initial and final position of each electron is recorded, and the number of electrons crossing each virtual plane is counted. Interpolation of the number of electrons as a function of virtual plane position with an exponential function allows for the calculation of the effective ionization Townsend coefficient, $\alpha_{eff}$, using equation \ref{eq:townsend}. To obtain $\alpha$, the interpolation is repeated without accounting for attachment processes. Finally, $\eta$ is calculated as the difference between $\alpha$ and $\alpha_{eff}$. Uncertainties of $\alpha$ and $\alpha_{eff}$ are assumed to be equal to the uncertainties of the corresponding best-fit functions on simulation data, and error propagation is used to determine the uncertainty of $\eta$.

The instantaneous mean electron energy $<\!\varepsilon(t)\!>$ and the instantaneous velocity $\vec{v}(t)$ of electrons in the avalanche are evaluated by implementing equations \ref{eq:energia_istantanea_MATOQ} and \ref{eq:cap5_formula3} where $K$ is here the number of electrons simulated at time $t$. Similarly to the temporal growth configuration, the mean electron energy $<\!\varepsilon\!>$ and the electron velocity $\vec{v}$ are calculated in MATOQ by averaging all respective values sampled at each time step $\Delta t$ after reaching the steady state.

\begin{figure}[ht]
    \centering
    \includegraphics[width=0.95\textwidth]{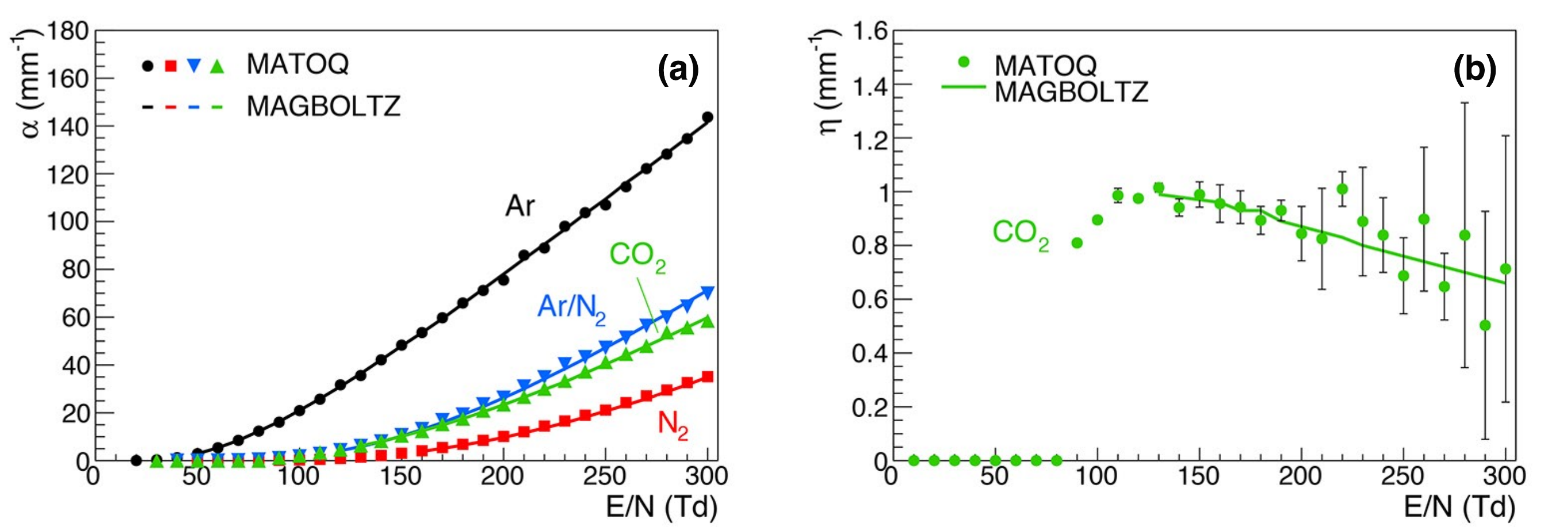}
    \caption{Values of ionization Townsend coefficient $\alpha$ (a) and attachment Townsend coefficient $\eta$ (b) as a function of the reduced electric field $E/N$ in pure Ar, N$_{2}$, CO$_{2}$ and in the gas mixture of 50\% Ar and 50\% N$_{2}$. Some statistical error bars are hidden by markers.}
    \label{fig:plot3}
\end{figure}

The validation of the spatial growth configuration of MATOQ is done by comparing the calculated values of $\alpha$ and $\eta$ as a function of $E/N$ with those obtained from MAGBOLTZ calculations in pure Ar, N$_{2}$, CO$_{2}$, and in the gas mixture of 50\% Ar and 50\% N$_{2}$. In the spatial growth configuration of MATOQ, the simulation ends after either reaching 1$\cdot$10$^{8}$ real collisions or simulating a maximum number of 10$^{6}$ electrons, whichever comes first. The steady state is assumed to be reached after 2.5$\cdot$10$^{7}$ real collisions, and 10 virtual planes are used to evaluate the values of $\alpha$ and $\eta$. Space charge effects are not taken into account in the MATOQ results to ensure consistency with the MAGBOLTZ calculations. The simulation is initiated with 300 electrons having an initial energy of 0.1 eV. Figure \ref{fig:plot3} shows that the values of $\alpha$ and $\eta$ obtained by MATOQ and MAGBOLTZ are in good agreement, with a relative error of less than $\sim$2\% for $\alpha$ and a maximum relative error of $\sim$25\% for $\eta$ from 10 Td to 300 Td.

\section{Spatial growth under the influence of the space charge electric field}\label{sec:spatial_space_charge}
The MATOQ program allows the simulation of the electron avalanche growth under the influence of an applied electric field together with the space charge electric field. As highlighted in section \ref{sec:space_charge}, electrons and ions are partially overlapped in space during the avalanche development in a gas volume. This causes the formation of the space charge electric field that is superimposed on the applied electric field. The externally applied electric field remains static and uniform in the gas medium, while the space charge electric field changes in space and time during the avalanche evolution. As a consequence, the applied electric field is strengthened in the upstream and downstream electron avalanche, whereas its strength results reduced in the middle of the avalanche. The presence of the space charge electric field reduces the electron multiplication in the gas volume compared to the case where it is absent \cite{abbrescia_book}.

To compute the space charge electric field in MATOQ, electron avalanche growth is simulated in a defined three-dimensional grid within a gas gap between two opposite electrodes. This approach contrasts with the simulation of electron avalanches in an infinite gas volume, as described in section \ref{sec:spatial}. The gas gap's width, the applied electric field strength, the initial electron positions, and the gas mixture's composition and volume fractions can be selected as desired. The spatial mesh, used to calculate the total electric field, can also be chosen. For simplicity, ions are assumed to be motionless, as their mobility is typically three orders of magnitude lower than that of electrons.

To compute the space charge effects during the growth of the avalanche, the gas gap is partitioned into cubic elements with position vectors $\vec{r}_{i}$ in a Cartesian coordinate system. Each grid point in the gas gap corresponds to a vector $\vec{r}_{i}$. The volume of each grid element is $\Delta x \cdot \Delta y \cdot \Delta z$. The applied electric field $\vec{E}$ is assumed to be parallel to the $z$-axis. Electron avalanches are initiated from a given number of initial electrons, placed anywhere in the gas gap. The simulation continues until all electrons reach the anode. The space charge electric field is computed for each grid point during the simulation and is recalculated after a given number of time steps $\Delta t$, arbitrarily selected by the user of MATOQ. This enables the dynamic evaluation of the space charge electric field for the entire duration of the electron avalanche.

The calculation of the electric field in the gas gap at a given time $t$ is carried out in four steps. Firstly, the total electric charge of each cubic element is calculated by counting how many electrons and ions are inside the corresponding cubic element of the grid. Secondly, the electric potential $V(\vec{r})$ in each grid point $\vec{r}_{i}$ is calculated at the time $t$ as:
\begin{equation}\label{eq:potenziale}
    V(\vec{r}_{i}) = \frac{1}{4 \pi \varepsilon} \sum_{a = 1}^{A} \frac{q(\vec{r_{a}})}{|\vec{r}_{i} - \vec{r_{a}}|}
\end{equation}
where $\varepsilon$ is the permittivity, $a$ indicates the $a$-th element in the spatial mesh, $A$ is the total number of cubic elements in which the gas gap is divided and, finally, $q(\vec{r_{a})}$ is the total electric charge in the cubic element identified by the position vector $\vec{r_{a}}$. The expression of the electric potential presents a singularity if $a = i$. In order to overcome this discontinuity, the difference $|\vec{r}_{i} - \vec{r_{a}}|$ is assumed equal to $\sqrt{(\Delta x/10)^{2} + (\Delta y/10)^{2} + (\Delta z/10)^{2}}$ in the case $\vec{r}_{i} = \vec{r}_{a}$. This means that the electric charge inside the $a$-th element is considered to be slightly displaced from the center of the cubic element where the electric potential is evaluated. In other words, the components of vector $\vec{r_{a}}$ along the $x$-, $y$-, and $z$-axes are respectively increased by a tenth of $\Delta x$, $\Delta y$ and $\Delta z$, which are the sizes of each cubic element. This is an arbitrary assumption that can be easily modified by the user of MATOQ as wished, however it gave satisfactory results in the calculation of avalanche sizes in a narrow-gap RPC, as will be demonstrated in the following. Thirdly, the total electric field at the time $t$ is calculated in each grid point $\vec{r}_{i}$ as the sum of the applied electric field $\vec{E}$ and the space charge electric field $\vec{E}_{sp-ch}(\vec{r}_{i})$, starting from the electric potential $V(\vec{r})$ in each grid point. Thirdly, the space charge electric field $\vec{E}_{sp-ch}(\vec{r}_{i})$ is calculated by evaluating the electric potential $V(\vec{r})$ in each grid point. Finally, the total electric field at time $t$ is computed as the sum of the applied electric field $\vec{E}$ and the space charge electric field $\vec{E}_{sp-ch}(\vec{r}_{i})$.

\begin{figure}[ht]
    \centering
    \includegraphics[width=0.50\textwidth]{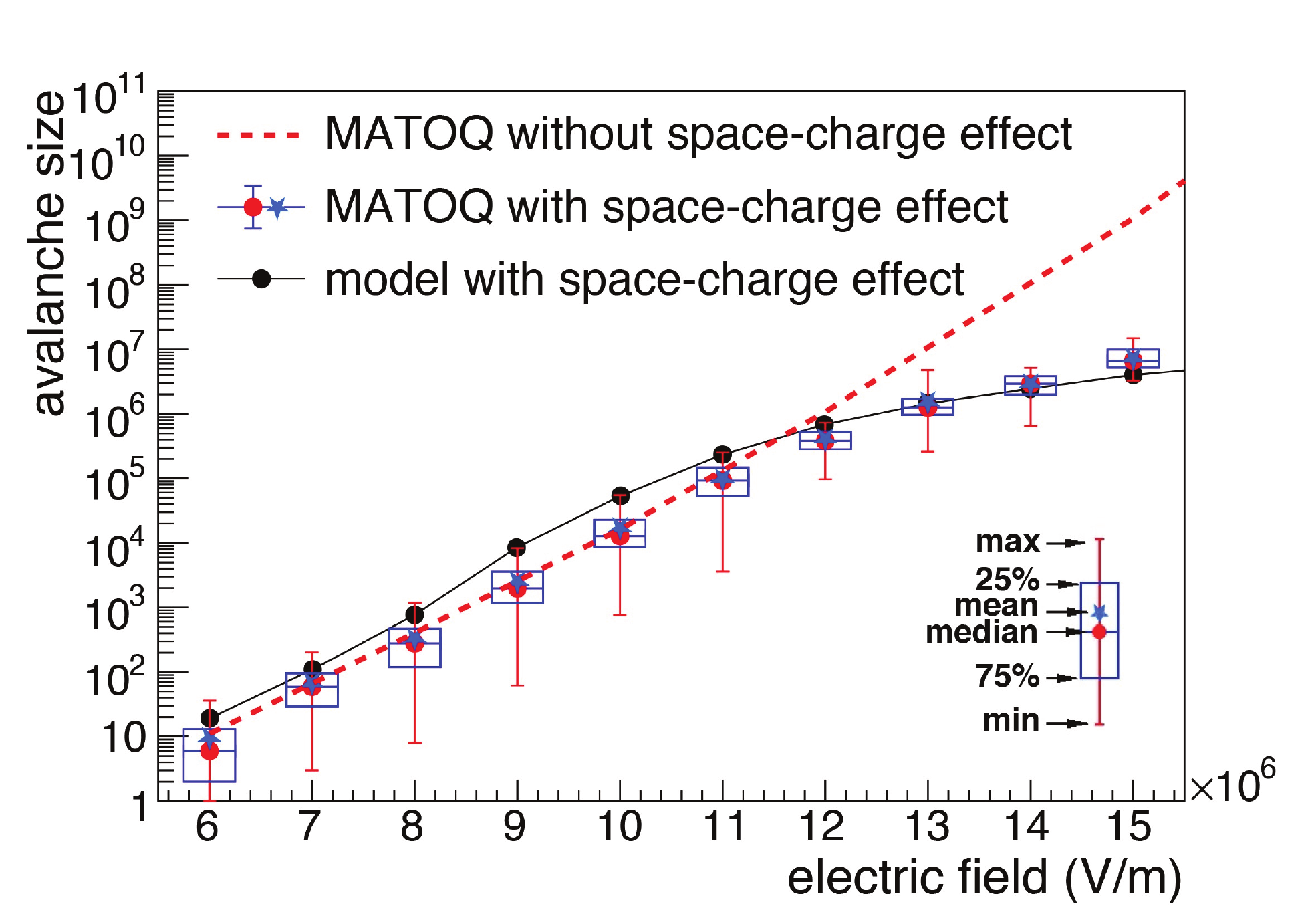}
    \caption{Avalanche sizes at the anode as a function of the electric field in a 0.1-mm single-gap RPC using a gas mixture of 85\% C$_{2}$H$_{2}$F$_{4}$, 5\% \textit{i}-C$_{4}$H$_{10}$ and 10\% SF$_{6}$ at 296.15 K and 970 mbar. Values of avalanche size obtained by the MATOQ simulation with and without considering the space charge effects are compared with the results of Lippmann et al.'s model, which takes into account those effects. Concerning the MATOQ results obtained by the simulation of space charge effects, the distribution of the avalanche sizes is represented by plotting the maximum and minimum value, the lower (25\%) and higher (75\%) quartile as well as the median and the mean value of the distribution. Data of the model are provided by Lippmann et al. in their paper \cite{lippmann2004space}.}
    \label{fig:plot6}
\end{figure}

Unlike previous cases in sections \ref{sec:temporal} and \ref{sec:spatial}, the MATOQ simulation in the spatial growth configuration under the influence of the space charge electric field cannot be validated by a comparison of results obtained by different simulation codes, like MAGBOLTZ or METHES. To validate the MATOQ simulations, the avalanche sizes at the anode as a function of the electric field are compared with values obtained by the Lippmann et al.'s model \cite{lippmann2004space} in a narrow-gap RPC. The size of the avalanche at the anode in this type of gaseous particle detector depends on certain parameters. The number and position of primary electrons, released by an incoming radiation that ionizes the gases in the detector, the gas mixture and its density, and the applied electric field play a crucial role in the charge multiplication in RPCs. Figure \ref{fig:plot6} shows the temporal evolution of the avalanche, simulated with MATOQ, for two different values of the applied electric field. All avalanches are originated from one single electron starting from the origin at $t$ = 0 s. The charge multiplication is simulated in a gas mixture of 85\% C$_{2}$H$_{2}$F$_{4}$, 5\% \textit{i}-C$_{4}$H$_{10}$ and 10\% SF$_{6}$ at 296.15 K and 970 mbar. The number of electrons in avalanches at 13 kV/mm increases in time faster than that at 9 kV/mm. For avalanches simulated at 13 kV/mm, the number of electrons in the gas gap is $\sim$10$^{6}$ electrons after $\sim$0.25 ns and then it progressively decreases until all electrons reach the anode, whereas avalanches at $\sim$9 kV/mm reach the maximum number of $\sim$10$^{3}$ electrons at $\sim$0.5 ns. This is caused by the fact that $\alpha_{eff}$ and $v_{drift}$ of the gas mixture at 13 kV/mm are higher than those at 9 kV/mm.
\begin{figure}[ht]
    \centering
    \includegraphics[width=0.50\textwidth]{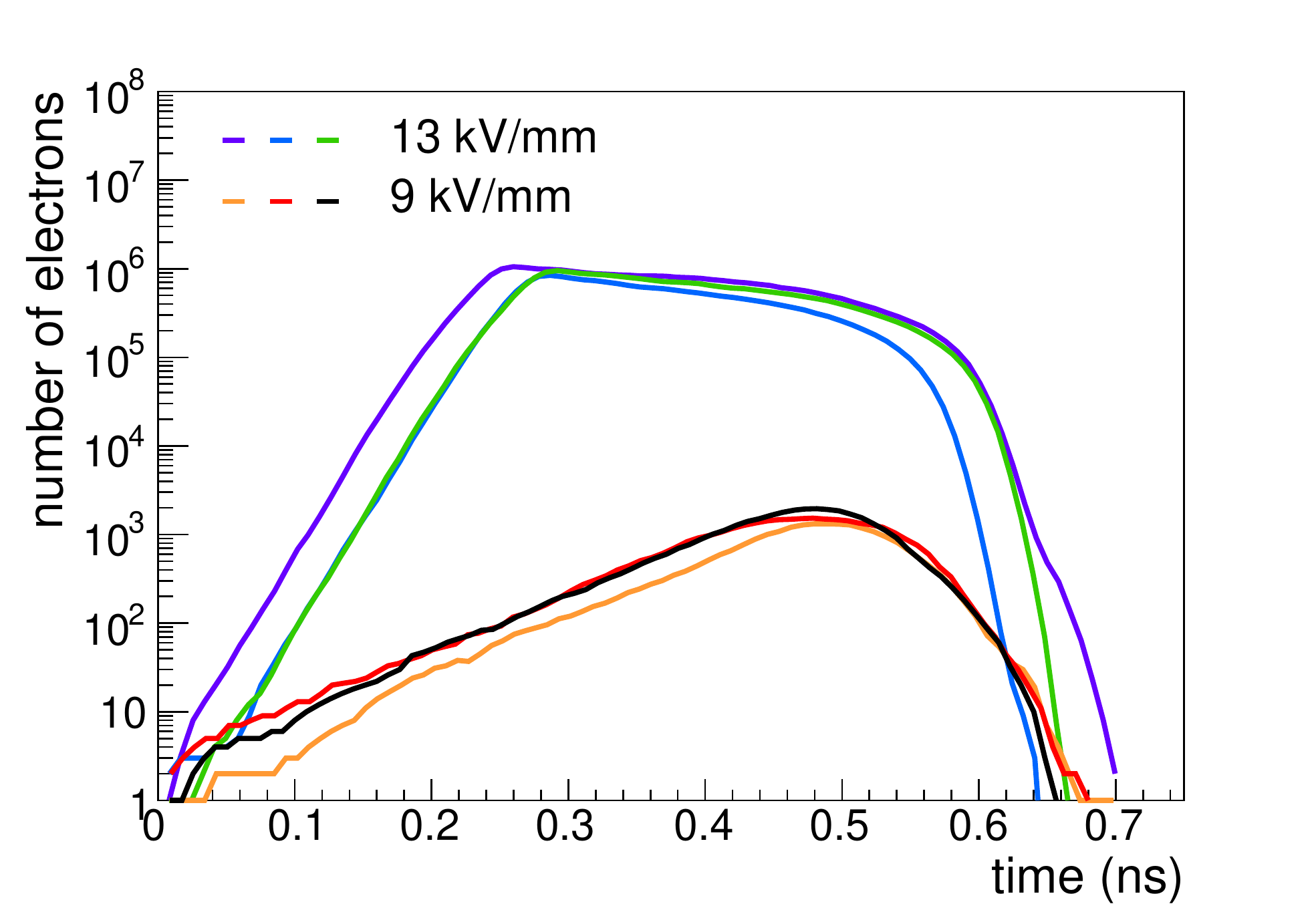}
    \caption{Number of electrons in six avalanches as a function of time in a gas gap of 0.1 mm at 9 kV/mm and 13 kV/mm. The MATOQ simulation results are obtained in the gas mixture of 85\% C$_{2}$H$_{2}$F$_{4}$, 5\% \textit{i}-C$_{4}$H$_{10}$ and 10\% SF$_{6}$ at 296.15 K and 970 mbar.}
    \label{fig:plot4}
\end{figure}

\begin{figure}[ht]
    \centering
    \includegraphics[width=0.95\textwidth]{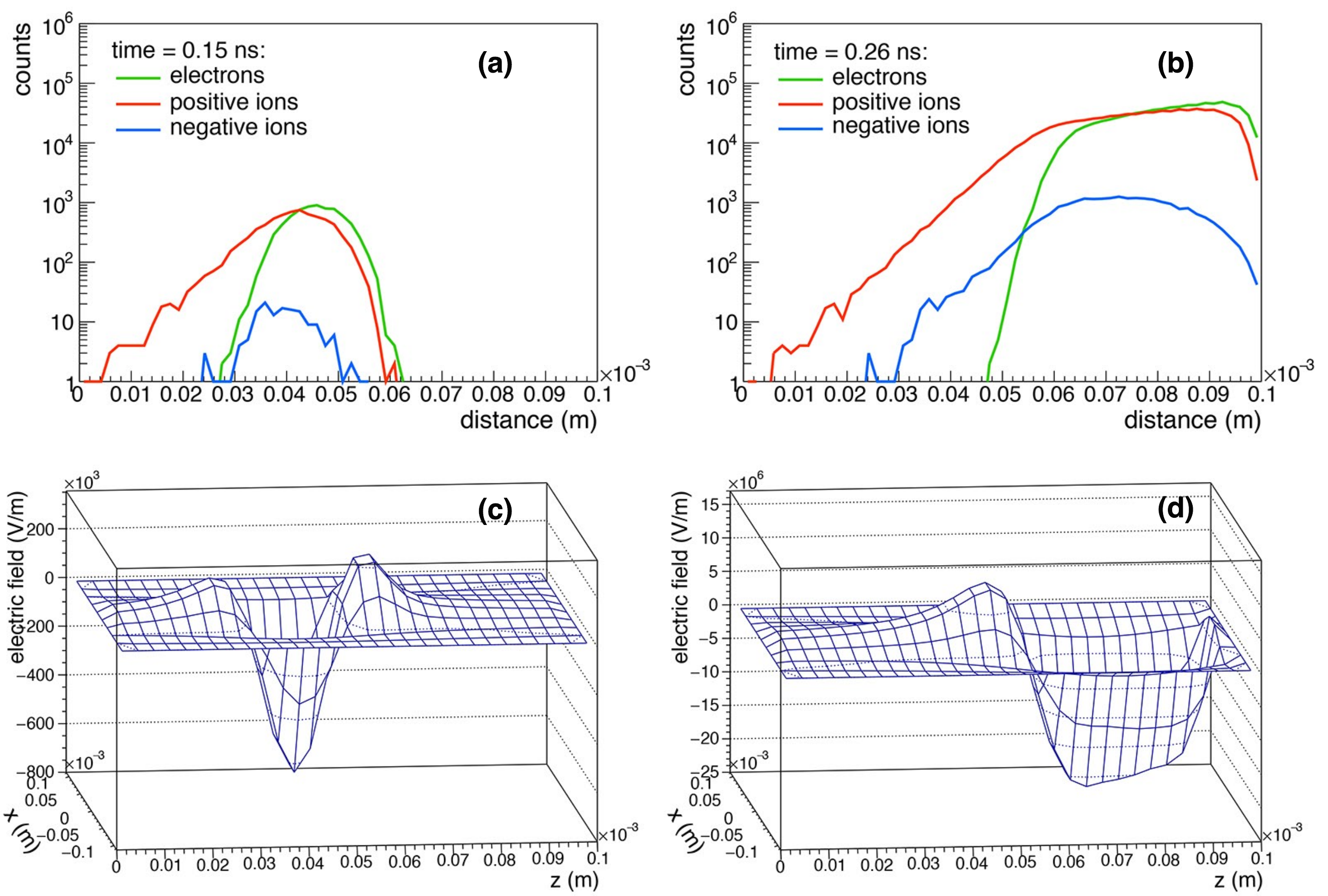}
    \caption{Number of electrons and ions as a function of the distance in the same avalanche at 14 kV/mm is shown at 0.15 ns (a) and 0.26 ns (b). Values of the space charge electric field are presented at 0.15 ns (c) and 0.26 ns (d) for the same avalanche. The MATOQ simulation is carried out in a 0.1-mm single-gap RPC using a gas mixture of 85\% C$_{2}$H$_{2}$F$_{4}$, 5\% \textit{i}-C$_{4}$H$_{10}$ and 10\% SF$_{6}$ at 296.15 K and 970 mbar.}
    \label{fig:plot5}
\end{figure}

Electrons and ions are partially overlapped in space during the avalanche growth in the gas gap. Figures \ref{fig:plot5}a and \ref{fig:plot5}b show the number of electrons and ions along the gas gap at 0.15 ns and 0.26 ns, respectively. In this case, the number of electrons and ions as a function of the distance are evaluated in an avalanche originated from one single electron, which starts from the origin and at $t$ = 0 s. The MATOQ simulation is carried out in the gas mixture of 85\% C$_{2}$H$_{2}$F$_{4}$, 5\% \textit{i}-C$_{4}$H$_{10}$ and 10\% SF$_{6}$ at 296.15 K and 970 mbar with an applied electric field of 14 kV/mm. The overlap of positive and negative charges generates the space charge electric field along the gas gap. Values of the space charge electric field at 0.15 ns and 0.26 ns are shown in figures \ref{fig:plot5}c and \ref{fig:plot5}d, respectively. The MATOQ simulation results in figure \ref{fig:plot5} are consistent with the findings of Lippmann et al.'s model \cite{lippmann2004space}. In particular, there are regions in the gas gap where the space charge electric field is decreased, while in other regions, it is increased. This effect becomes more evident during the evolution of the avalanche in both space and time.

A comparison between avalanche sizes at the anode simulated with MATOQ and those calculated with the Lippmann et al.'s model is presented in figure \ref{fig:plot6}. The comparison is carried out in a 0.1-mm single-gap RPC, using a gas mixture of 85\% C$_{2}$H$_{2}$F$_{4}$, 5\% \textit{i}-C$_{4}$H$_{10}$ and 10\% SF$_{6}$ at 296.15 K and 970 mbar, for electric fields ranging from 6 kV/mm to 15 kV/mm. In addition, the average size of avalanches originating from a single electron at the cathode is calculated in the absence of space charge effects and shown in figure \ref{fig:plot6}. The appearance of space charge effects during the avalanche development generally leads to a decrease in gas gain and reduced avalanche sizes \cite{abbrescia_book}. The agreement between the average sizes obtained from MATOQ simulations, with and without considering the space charge effects, is good for low electric field values. On the contrary, for electric field values higher than 12 kV/mm, the difference between the average avalanche size calculated with and without space charge effects becomes significant. Specifically, at 15 kV/mm, the average avalanche size without considering space charge effects is two orders of magnitude higher than that obtained by accounting for the space charge effects. Figure \ref{fig:plot6} also shows the average avalanche sizes computed using the Lippmann et al.'s model. The MATOQ simulation results with space charge effects exhibit good agreement with the calculations performed by Lippmann et al., except for electric field values below 11 kV/mm where there is a difference by a factor of approximately 2. There could be some reasons for the discrepancies between the results of the Lippmann et al.'s model and the MATOQ simulation. Firstly, the Lippmann et al.'s model generates avalanches using a pattern of initial electrons with an accurate estimation of their positions and energies, while all avalanches in MATOQ are generated by a single electron at the cathode with an initial energy of 5 eV. Secondly, there could be differences in the electron collision cross sections used in the model and in the simulation. Lippmann et al. used MAGBOLTZ 2.2 to evaluate electron transport coefficients and reaction rates, while MATOQ uses electron collision cross sections from different databases. In fact, electron collision cross sections of SF$_{6}$ and \textit{i}-C$_{4}$H$_{10}$ for the MATOQ simulation are the same as those implemented in MAGBOLTZ 10.6, whereas the cross sections for C$_{2}$H$_{2}$F$_{4}$ are provided by {\v{S}}a{\v{s}}i{\'c} et al. in their work published in 2013 \cite{vsavsicswarm}.

During testing the MATOQ code, a limitation has been found in the simulation of more than $\sim$5$\cdot$10$^{7}$ electrons. Indeed, several simulations of avalanches with this large number of electrons turned out to be incomplete probably because of a memory allocation limitation. The amount of simulation data to temporarily record might have saturated the available memory of the platform where the code was running. Nevertheless, this does not represent an important limitation if the gas gain is not too high, as in the case shown in figure \ref{fig:plot6}. More details of the system where the MATOQ program is executed are provided in the appendix. 

\section{Electron Transport Parameters in C$_{3}$H$_{2}$F$_{4}$-based Gas Mixtures}\label{sec:electron_transport}
The reduction of fluorinated greenhouse gas emissions in the European Union countries has been made mandatory by new regulations \cite{europeanparliament} introduced since January 2015. The primary objective of the regulation is to gradually phase out hydrofluorocarbons (such as C$_{2}$H$_{2}$F$_{4}$), currently available on the market, to limit their overall production. Even though research applications are exempt from current regulations, the phasing out of hydrofluorocarbons could gradually increase their price due to their limited future availability. A number of R\&D studies \cite{abbrescia2016eco, guida2016characterization, liberti2016further, bianchi2019characterization} are ongoing to investigate the potential replacement of C$_{2}$H$_{2}$F$_{4}$-based gas mixtures for RPCs with other more environmental-friendly gases. One alternative to C$_{2}$H$_{2}$F$_{4}$ is C$_{3}$H$_{2}$F$_{4}$, which appears to be a viable solution for RPCs. However, directly replacing C$_{2}$H$_{2}$F$_{4}$ with C$_{3}$H$_{2}$F$_{4}$ is not feasible due to the resulting high operating voltages of RPCs. A potential solution to address this issue is to replace C$_{2}$H$_{2}$F$_{4}$ with a binary mixture of C$_{2}$H$_{2}$F$_{4}$ and CO$_{2}$ in varying proportions. 

\begin{figure}[ht]
    \centering
    \includegraphics[width=0.95\textwidth]{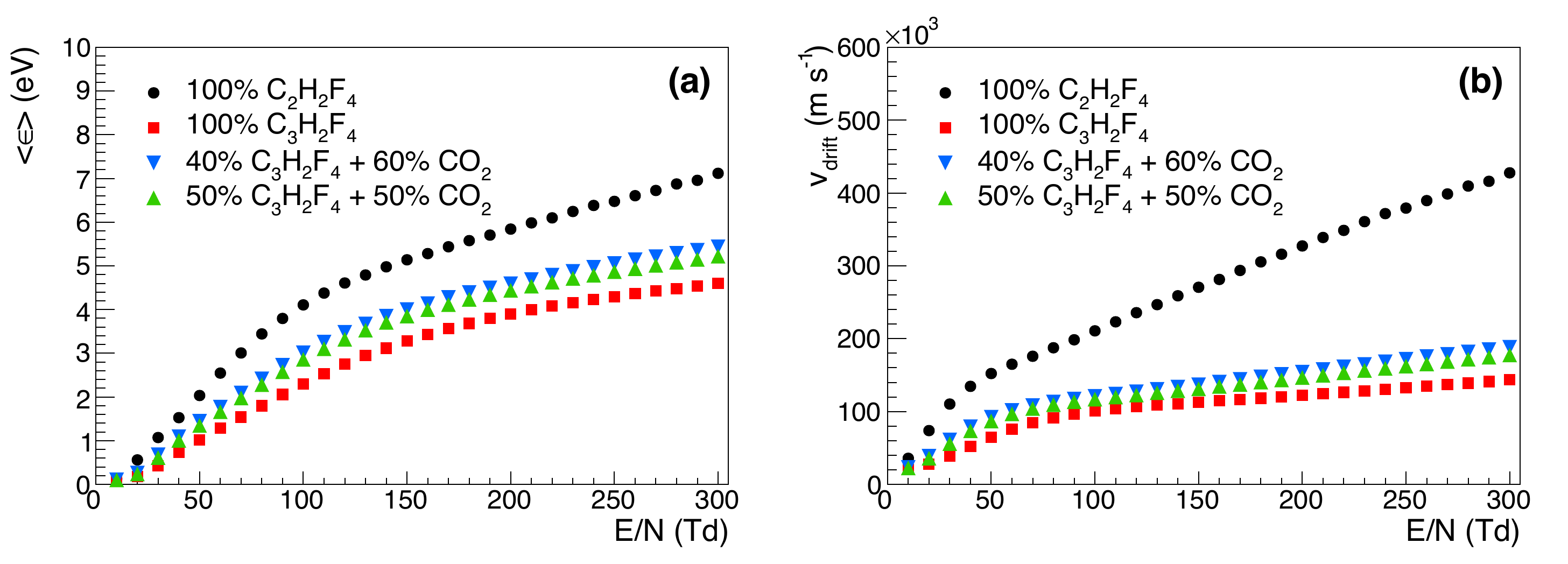}
    \caption{Values of average electron energy $<\!\varepsilon\!>$ (a) and drift velocity $v_{drift}$ (b) as a function of the reduced electric field $E/N$ in pure C$_{2}$H$_{2}$F$_{4}$, C$_{3}$H$_{2}$F$_{4}$, and in the C$_{3}$H$_{2}$F$_{4}$-based gas mixtures with 50\% or 60\% CO$_{2}$. Some statistical error bars are hidden by markers.}
    \label{fig:plot7}
\end{figure}

By using the MATOQ code, we can compare how the electron transport parameters vary in C$_{2}$H$_{2}$F$_{4}$ and C$_{3}$H$_{2}$F$_{4}$ as well as in gas mixtures of C$_{3}$H$_{2}$F$_{4}$ and CO$_{2}$ in different percentages. Figures \ref{fig:plot7}a and \ref{fig:plot7}b show the $<\!\varepsilon\!>$ and $v_{drift}$ in C$_{2}$H$_{2}$F$_{4}$ and C$_{3}$H$_{2}$F$_{4}$ as a function of the reduced electric field $E/N$, respectively. In pure C$_{2}$H$_{2}$F$_{4}$, values of $<\!\varepsilon\!>$ and $v_{drift}$ are higher than those in pure C$_{3}$H$_{2}$F$_{4}$ between 10 Td and 300 Td. At a reduced electric field of 150 Td, the average electron energy $<\!\varepsilon\!>$ is $\sim$5 eV in pure C$_{2}$H$_{2}$F$_{4}$, while it is $\sim$3 eV in pure C$_{3}$H$_{2}$F$_{4}$. These values are $\sim$7 eV and $\sim$4.5 eV at 300 Td. Similarly, the drift velocity $v_{drift}$ in C$_{3}$H$_{2}$F$_{4}$ at 150 Td is about three times lower than that in C$_{2}$H$_{2}$F$_{4}$, with the reduction being about four times at 300 Td. When C$_{3}$H$_{2}$F$_{4}$ is mixed with 50\% or 60\% of CO$_{2}$, the values of $<\!\varepsilon\!>$ and $v_{drift}$ increase slightly as a function of $E/N$, compared to those in pure C$_{3}$H$_{2}$F$_{4}$, as shown in figures \ref{fig:plot7}a and \ref{fig:plot7}b. The increase of $<\!\varepsilon\!>$ is $\sim$20\% at both 150 Td and 300 Td. Similarly, $v_{drift}$ in C$_{3}$H$_{2}$F$_{4}$/CO$_{2}$ mixtures is increased by approximately 20\% at 150 Td compared to that in pure C$_{3}$H$_{2}$F$_{4}$, whereas the increase is $\sim$25\% at 300 Td.

\begin{figure}[ht]
    \centering
    \includegraphics[width=0.95\textwidth]{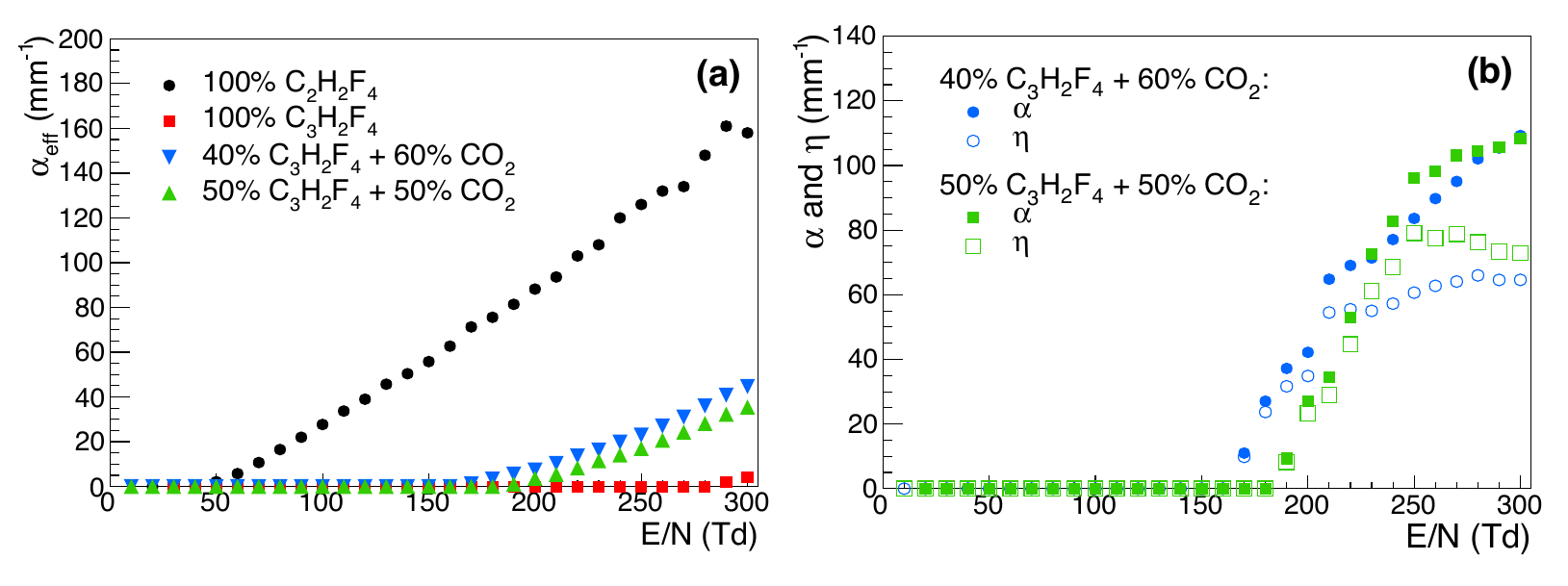}
    \caption{(a) Effective ionization Townsend coefficients $\alpha_{eff}$ as a function of $E/N$ in pure C$_{2}$H$_{2}$F$_{4}$, C$_{3}$H$_{2}$F$_{4}$, and  in the C$_{3}$H$_{2}$F$_{4}$-based gas mixtures with 50\% and 60\% CO$_{2}$. (b) Ionization Townsend coefficient $\alpha$ and attachment Townsend coefficient $\eta$ as a function of $E/N$ in C$_{3}$H$_{2}$F$_{4}$-based gas mixtures with CO$_{2}$. Some statistical error bars are hidden by markers.}
    \label{fig:plot8}
\end{figure}

Figure \ref{fig:plot8}a shows the positive values of the effective ionization Townsend coefficient $\alpha_{eff}$ as a function of $E/N$ in pure C$_{2}$H$_{2}$F$_{4}$, pure C$_{3}$H$_{2}$F$_{4}$, and C$_{3}$H$_{2}$F$_{4}$-based gas mixtures with 50\% or 60\% CO$_{2}$. These values are reported to identify the reduced electric field values at which electron avalanches can occur in RPCs. For pure C$_{2}$H$_{2}$F$_{4}$, the effective ionization Townsend coefficient is higher than 0, indicating that ionization events occur more frequently than attachment events, at a reduced electric field of $\sim$50 Td. On the contrary, the electron avalanche growth can occur at $\sim$290 Td in pure C$_{3}$H$_{2}$F$_{4}$. When C$_{3}$H$_{2}$F$_{4}$ is diluted with CO$_{2}$, the growth of electron avalanches occurs at lower values of the electric field, as reported by experimental studies \cite{bianchi2019characterization, proto2022new, abbrescia2016preliminary, rigoletti2020studies, bianchi2021electron}. This observation is also supported by MATOQ simulations. Indeed, in a gas mixture with equal proportions of C$_{3}$H$_{2}$F$_{4}$ and CO$_{2}$, $\alpha_{eff}$ begins to exceed 0 at $\sim$190 Td. If the CO$_{2}$ percentage is increased from 50\% to 60\%, ionization events are more frequent than attachment events starting from 170 Td. Figure \ref{fig:plot8}b shows the ionization Townsend coefficient $\alpha$ and the attachment Townsend coefficient $\eta$ as a function of $E/N$ in C$_{3}$H$_{2}$F$_{4}$-based gas mixture with the addition of 50\% of CO$_{2}$ or 60\% of CO$_{2}$. As shown in figure \ref{fig:plot8}b, values of $\alpha$ increase progressively with $E/N$, whereas values of $\eta$ tend to reach a plateau after the initial growth.

\begin{figure}[ht]
    \centering
    \includegraphics[width=0.50\textwidth]{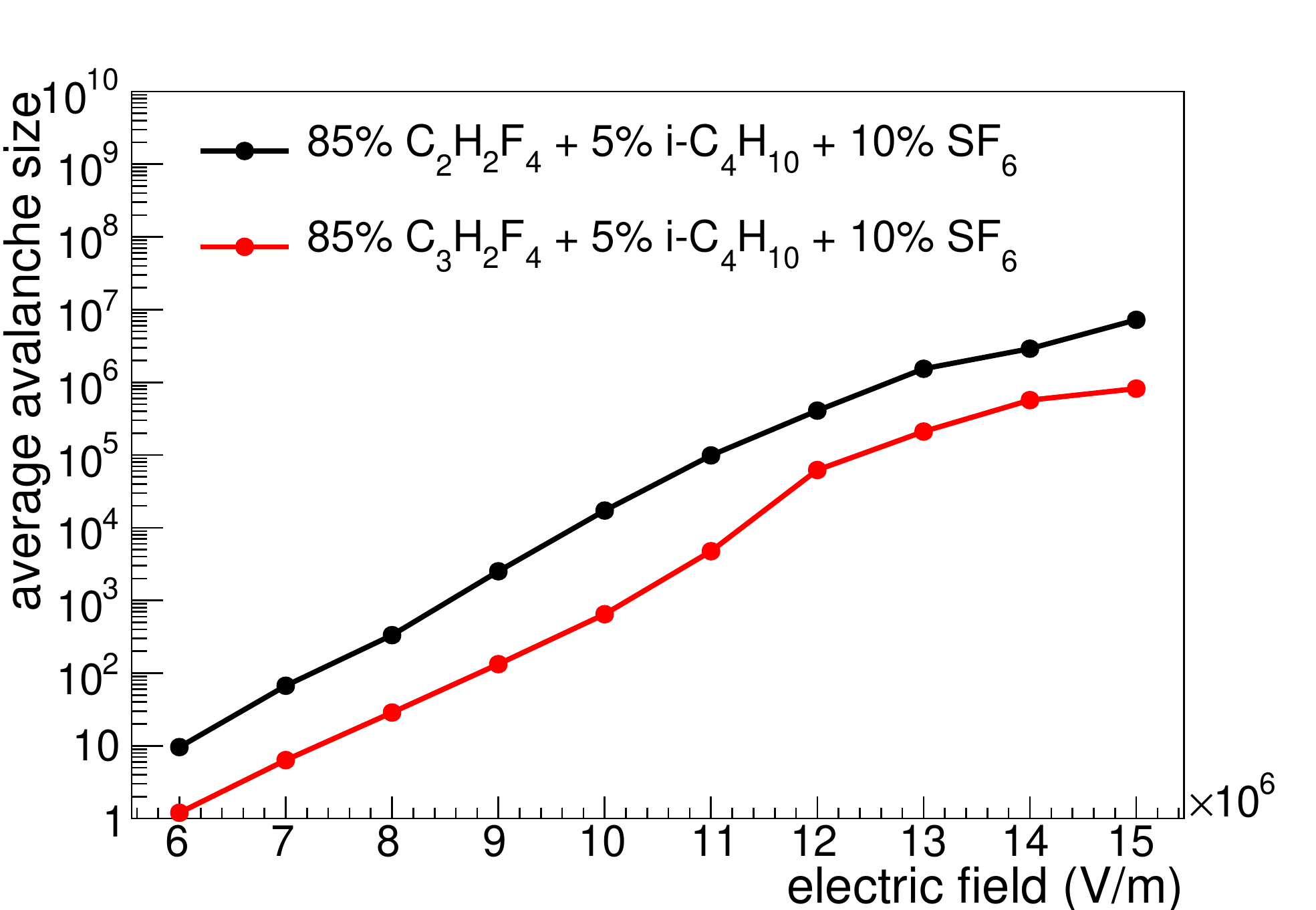}
    \caption{Average avalanche size at the anode in an RPC with a gas gap of 0.1 mm ranging from 6 kV/mm to 15 kV/mm at 296.15 K and 970 mbar. The MATOQ simulation is conducted using a C$_{2}$H$_{2}$F$_{4}$-based gas mixture (black) and a C$_{3}$H$_{2}$F$_{4}$-based gas mixture (red), both with the addition of 5\% \textit{i}-C$_{4}$H$_{10}$ and 10\% SF$_{6}$. Simulations consider the space charge effects in the gas gap. Some statistical error bars are hidden by markers.}
    \label{fig:plot9}
\end{figure}

The replacement of C$_{2}$H$_{2}$F$_{4}$ with C$_{3}$H$_{2}$F$_{4}$ requires dedicated studies also in narrow-gap RPCs, where the effect of space charge electric field may play a crucial role. Figure \ref{fig:plot9} presents the average avalanche size at the anode in a 0.1-mm single-gap RPC. The simulation is carried out for a range of applied electric fields from 6 kV/mm to 15 kV/mm, including the space charge effects in the gas gap and using the gas mixture consisting of 85\% C$_{2}$H$_{2}$F$_{4}$, 5\% \textit{i}-C$_{4}$H$_{10}$ and 10\% SF$_{6}$ at 296.15 K and 970 mbar. Additionally, the simulations are performed by replacing C$_{2}$H$_{2}$F$_{4}$ with an equal amount of C$_{3}$H$_{2}$F$_{4}$. In the C$_{3}$H$_{2}$F$_{4}$-based gas mixture, the average avalanche size follows an exponential trend until approximately 10 kV/mm. At low electric field values, the average avalanche size is reduced by a factor of 20 to 30 when C$_{2}$H$_{2}$F$_{4}$ is replaced with C$_{3}$H$_{2}$F$_{4}$. On the contrary, the average avalanche size in the C$_{3}$H$_{2}$F$_{4}$-based gas mixture is approximately 5$-$8 times smaller than that in the C$_{2}$H$_{2}$F$_{4}$-based gas mixture at high electric field values.

\section{Conclusions}\label{sec:conclusion}
The MATOQ program has been developed to study environmental-friendly gas mixtures for RPCs. This program enables the simulation of electron transport coefficients and reaction rates in gases under the influence of an electric field. Mean energy and drift velocity of electrons as well as ionization and attachment coefficients are evaluated both in temporal and in spatial growth configurations. Unlike the already existing programs, the MATOQ code also allows the evaluation of space charge effects between electrons and ions during the development of the electron avalanche. MATOQ is written in C++, which makes it a multi-platform software, and supports the multi-thread execution, which speeds the computation time. The data format of electron collision cross sections adopted in the LXCat database, which is widely used and regularly upgraded, is compatible with MATOQ.

The temporal and spatial growth configurations of electron avalanches under the influence of uniform electric fields have been validated by comparing the MATOQ results with those obtained by MAGBOLTZ. The electron transport coefficients and reaction rates, namely $\varepsilon$, $v_{drift}$, $\nu_{ion}$, $\nu_{att}$, $\alpha$ and $\eta$, show good agreement with experimental data in pure Ar, N$_{2}$, CO$_{2}$ and in the binary mixture of 50\% Ar and 50\% N$_{2}$. Moreover, the simulation of electron avalanches influenced by both uniform and space charge electric fields is validated by comparing the avalanche sizes obtained from MATOQ to those calculated using Lippmann et al.'s model.

According to several experimental R\&D studies, one potential alternative to C$_{2}$H$_{2}$F$_{4}$ in RPCs is C$_{3}$H$_{2}$F$_{4}$, which is considered to be more environmentally friendly. Using MATOQ, we calculated the changes in electron transport coefficients and reaction rates in pure C$_{2}$H$_{2}$F$_{4}$ and C$_{3}$H$_{2}$F$_{4}$, as well as in gas mixtures of C$_{3}$H$_{2}$F$_{4}$ and CO$_{2}$ in various proportions. The dilution of C$_{3}$H$_{2}$F$_{4}$ with CO$_{2}$ is considered a viable solution for operating RPCs within the voltage range currently used. This solution may also be applicable to narrow-gap RPCs. Indeed, the simulations conducted using MATOQ suggest that the average avalanche size is reduced approximately by one order of magnitude when C$_{2}$H$_{2}$F$_{4}$ is only replaced with C$_{3}$H$_{2}$F$_{4}$ in a 0.1-mm single-gap RPC.

\section*{Appendix}
All sets of electron collision cross sections used in this work as input for MATOQ are summarized in table \ref{tab:tabella_sezioni}:
\begin{table}[ht]
\begin{center}
 \begin{tabular}{c c} 
 \hline
 Gas & Source \\ 
 \hline\hline
 Ar & LXCat: Biagi's database MAGBOLTZ v8.97 \cite{lxcat_database}  \\ 
 \hline
 N$_{2}$ & LXCat: Biagi's database MAGBOLTZ v8.97 \cite{lxcat_database} \\
 \hline
 CO$_{2}$ & LXCat: Biagi's database MAGBOLTZ v11.6 \cite{lxcat_database} \\
 \hline
 C$_{2}$H$_{2}$F$_{4}$ & {\v{S}}a{\v{s}}i{\'c} et al. \cite{vsavsicswarm} \\
 \hline
  C$_{3}$H$_{2}$F$_{4}$ & Bianchi et al. \cite{bianchi2021electron} \\
 \hline
  \textit{i}-C$_{4}$H$_{10}$ & MAGBOLTZ v10.6 \cite{biagi1999monte}\\ 
 \hline
 SF$_{6}$ & LXCat: Biagi's database MAGBOLTZ v10.6 \cite{lxcat_database} \\ 
 \hline
\end{tabular}
\caption{Sets of electron collision cross sections used for the MATOQ simulations of this work.}
\label{tab:tabella_sezioni}
\end{center}
\end{table}

Comparisons between MATOQ and MAGBOLTZ calculations in figures \ref{fig:plot2} and \ref{fig:plot3} are carried out by using the version 8.97 of MAGBOLTZ in the case of Ar and N$_{2}$, whereas the version 11.6 of MAGBOLTZ is used for the result simulations in pure CO$_{2}$, according to table \ref{tab:tabella_sezioni}.

The MATOQ program is interfaced with the software ROOT, freely provided by the European Organization for Nuclear Research (CERN), to plot the results during the simulation \cite{root_documentation}.

All simulations of this work have been performed in the virtual machines of the Linux Public Login User Service (LXPLUX7) provided by CERN. These machines are organized in a cluster of PCs running CERN CentOS Linux in 64-bit mode. More details can be found in the LXPLUS7 documentation.

\section*{\label{sec:level14}DATA AVAILABILITY}
The data that support the findings of this study are available from the corresponding author upon reasonable request.

\nocite{*}

\bibliography{apssamp}

\begin{thebibliography}{31}%
\makeatletter
\providecommand \@ifxundefined [1]{%
 \@ifx{#1\undefined}
}%
\providecommand \@ifnum [1]{%
 \ifnum #1\expandafter \@firstoftwo
 \else \expandafter \@secondoftwo
 \fi
}%
\providecommand \@ifx [1]{%
 \ifx #1\expandafter \@firstoftwo
 \else \expandafter \@secondoftwo
 \fi
}%
\providecommand \natexlab [1]{#1}%
\providecommand \enquote  [1]{``#1''}%
\providecommand \bibnamefont  [1]{#1}%
\providecommand \bibfnamefont [1]{#1}%
\providecommand \citenamefont [1]{#1}%
\providecommand \href@noop [0]{\@secondoftwo}%
\providecommand \href [0]{\begingroup \@sanitize@url \@href}%
\providecommand \@href[1]{\@@startlink{#1}\@@href}%
\providecommand \@@href[1]{\endgroup#1\@@endlink}%
\providecommand \@sanitize@url [0]{\catcode `\\12\catcode `\$12\catcode
  `\&12\catcode `\#12\catcode `\^12\catcode `\_12\catcode `\%12\relax}%
\providecommand \@@startlink[1]{}%
\providecommand \@@endlink[0]{}%
\providecommand \url  [0]{\begingroup\@sanitize@url \@url }%
\providecommand \@url [1]{\endgroup\@href {#1}{\urlprefix }}%
\providecommand \urlprefix  [0]{URL }%
\providecommand \Eprint [0]{\href }%
\providecommand \doibase [0]{https://doi.org/}%
\providecommand \selectlanguage [0]{\@gobble}%
\providecommand \bibinfo  [0]{\@secondoftwo}%
\providecommand \bibfield  [0]{\@secondoftwo}%
\providecommand \translation [1]{[#1]}%
\providecommand \BibitemOpen [0]{}%
\providecommand \bibitemStop [0]{}%
\providecommand \bibitemNoStop [0]{.\EOS\space}%
\providecommand \EOS [0]{\spacefactor3000\relax}%
\providecommand \BibitemShut  [1]{\csname bibitem#1\endcsname}%
\let\auto@bib@innerbib\@empty
\bibitem [{\citenamefont {Santonico}\ and\ \citenamefont
  {Cardarelli}(1981)}]{santonico1981development}%
  \BibitemOpen
  \bibfield  {author} {\bibinfo {author} {\bibfnamefont {R.}~\bibnamefont
  {Santonico}}\ and\ \bibinfo {author} {\bibfnamefont {R.}~\bibnamefont
  {Cardarelli}},\ }\bibfield  {title} {\bibinfo {title} {Development of
  {R}esistive {P}late {C}ounters},\ }\href@noop {} {\bibfield  {journal}
  {\bibinfo  {journal} {Nuclear Instruments and Methods in physics research}\
  }\textbf {\bibinfo {volume} {187}},\ \bibinfo {pages} {377} (\bibinfo {year}
  {1981})}\BibitemShut {NoStop}%
\bibitem [{\citenamefont {Bruno}(2004)}]{bruno2004resistive}%
  \BibitemOpen
  \bibfield  {author} {\bibinfo {author} {\bibfnamefont {G.}~\bibnamefont
  {Bruno}},\ }\bibfield  {title} {\bibinfo {title} {{R}esistive {P}late
  {C}hambers in running and future experiments},\ }\href@noop {} {\bibfield
  {journal} {\bibinfo  {journal} {The European Physical Journal C-Particles and
  Fields}\ }\textbf {\bibinfo {volume} {33}},\ \bibinfo {pages} {s1032}
  (\bibinfo {year} {2004})}\BibitemShut {NoStop}%
\bibitem [{\citenamefont {Abbrescia}\ \emph {et~al.}(2012)\citenamefont
  {Abbrescia}, \citenamefont {Aiola}, \citenamefont {Antolini}, \citenamefont
  {Avanzini}, \citenamefont {Ferroli}, \citenamefont {Bencivenni},
  \citenamefont {Bossini}, \citenamefont {Bressan}, \citenamefont {Chiavassa},
  \citenamefont {Cicalo} \emph {et~al.}}]{abbrescia2012eee}%
  \BibitemOpen
  \bibfield  {author} {\bibinfo {author} {\bibfnamefont {M.}~\bibnamefont
  {Abbrescia}}, \bibinfo {author} {\bibfnamefont {S.}~\bibnamefont {Aiola}},
  \bibinfo {author} {\bibfnamefont {R.}~\bibnamefont {Antolini}}, \bibinfo
  {author} {\bibfnamefont {C.}~\bibnamefont {Avanzini}}, \bibinfo {author}
  {\bibfnamefont {R.~B.}\ \bibnamefont {Ferroli}}, \bibinfo {author}
  {\bibfnamefont {G.}~\bibnamefont {Bencivenni}}, \bibinfo {author}
  {\bibfnamefont {E.}~\bibnamefont {Bossini}}, \bibinfo {author} {\bibfnamefont
  {E.}~\bibnamefont {Bressan}}, \bibinfo {author} {\bibfnamefont
  {A.}~\bibnamefont {Chiavassa}}, \bibinfo {author} {\bibfnamefont
  {C.}~\bibnamefont {Cicalo}}, \emph {et~al.},\ }\bibfield  {title} {\bibinfo
  {title} {The {EEE} project: cosmic rays, multigap {R}esistive {P}late
  {C}hambers and high school students},\ }\href@noop {} {\bibfield  {journal}
  {\bibinfo  {journal} {Journal of Instrumentation}\ }\textbf {\bibinfo
  {volume} {7}}\bibinfo  {number} { (11)},\ \bibinfo {pages}
  {P11011}}\BibitemShut {NoStop}%
\bibitem [{\citenamefont {Zeballos}\ \emph
  {et~al.}(1996{\natexlab{a}})\citenamefont {Zeballos}, \citenamefont {Crotty},
  \citenamefont {Hatzifotiadou}, \citenamefont {Valverde}, \citenamefont
  {Neupane}, \citenamefont {Williams},\ and\ \citenamefont
  {Zichichi}}]{zeballos1996new}%
  \BibitemOpen
\bibfield  {number} {  }\bibfield  {author} {\bibinfo {author} {\bibfnamefont
  {E.~C.}\ \bibnamefont {Zeballos}}, \bibinfo {author} {\bibfnamefont
  {I.}~\bibnamefont {Crotty}}, \bibinfo {author} {\bibfnamefont
  {D.}~\bibnamefont {Hatzifotiadou}}, \bibinfo {author} {\bibfnamefont {J.~L.}\
  \bibnamefont {Valverde}}, \bibinfo {author} {\bibfnamefont {S.}~\bibnamefont
  {Neupane}}, \bibinfo {author} {\bibfnamefont {M.}~\bibnamefont {Williams}},\
  and\ \bibinfo {author} {\bibfnamefont {A.}~\bibnamefont {Zichichi}},\
  }\bibfield  {title} {\bibinfo {title} {A new type of {R}esistive {P}late
  {C}hamber: The {M}ultigap {RPC}},\ }\href@noop {} {\bibfield  {journal}
  {\bibinfo  {journal} {Nuclear Instruments and Methods in Physics Research
  Section A: Accelerators, Spectrometers, Detectors and Associated Equipment}\
  }\textbf {\bibinfo {volume} {374}},\ \bibinfo {pages} {132} (\bibinfo {year}
  {1996}{\natexlab{a}})}\BibitemShut {NoStop}%
\bibitem [{\citenamefont {Amaldi}\ \emph {et~al.}(2015)\citenamefont {Amaldi},
  \citenamefont {Borghi}, \citenamefont {Bucciantonio}, \citenamefont
  {Kieffer}, \citenamefont {Samarati}, \citenamefont {Sauli},\ and\
  \citenamefont {Watts}}]{amaldi2015development}%
  \BibitemOpen
  \bibfield  {author} {\bibinfo {author} {\bibfnamefont {U.}~\bibnamefont
  {Amaldi}}, \bibinfo {author} {\bibfnamefont {G.}~\bibnamefont {Borghi}},
  \bibinfo {author} {\bibfnamefont {M.}~\bibnamefont {Bucciantonio}}, \bibinfo
  {author} {\bibfnamefont {R.}~\bibnamefont {Kieffer}}, \bibinfo {author}
  {\bibfnamefont {J.}~\bibnamefont {Samarati}}, \bibinfo {author}
  {\bibfnamefont {F.}~\bibnamefont {Sauli}},\ and\ \bibinfo {author}
  {\bibfnamefont {D.}~\bibnamefont {Watts}},\ }\bibfield  {title} {\bibinfo
  {title} {Development of {TOF}-{PET} detectors based on the multi-gap
  {R}esistive {P}late {C}hambers},\ }\href@noop {} {\bibfield  {journal}
  {\bibinfo  {journal} {Nuclear Instruments and Methods in Physics Research
  Section A: Accelerators, Spectrometers, Detectors and Associated Equipment}\
  }\textbf {\bibinfo {volume} {778}},\ \bibinfo {pages} {85} (\bibinfo {year}
  {2015})}\BibitemShut {NoStop}%
\bibitem [{\citenamefont {Crespo}\ \emph {et~al.}(2013)\citenamefont {Crespo},
  \citenamefont {Blanco}, \citenamefont {Couceiro}, \citenamefont {Ferreira},
  \citenamefont {Lopes}, \citenamefont {Martins}, \citenamefont
  {Ferreira~Marques},\ and\ \citenamefont {Fonte}}]{crespo2013resistive}%
  \BibitemOpen
  \bibfield  {author} {\bibinfo {author} {\bibfnamefont {P.}~\bibnamefont
  {Crespo}}, \bibinfo {author} {\bibfnamefont {A.}~\bibnamefont {Blanco}},
  \bibinfo {author} {\bibfnamefont {M.}~\bibnamefont {Couceiro}}, \bibinfo
  {author} {\bibfnamefont {N.~C.}\ \bibnamefont {Ferreira}}, \bibinfo {author}
  {\bibfnamefont {L.}~\bibnamefont {Lopes}}, \bibinfo {author} {\bibfnamefont
  {P.}~\bibnamefont {Martins}}, \bibinfo {author} {\bibfnamefont
  {R.}~\bibnamefont {Ferreira~Marques}},\ and\ \bibinfo {author} {\bibfnamefont
  {P.}~\bibnamefont {Fonte}},\ }\bibfield  {title} {\bibinfo {title}
  {{R}esistive {P}late {C}hambers in positron emission tomography},\
  }\href@noop {} {\bibfield  {journal} {\bibinfo  {journal} {The European
  Physical Journal Plus}\ }\textbf {\bibinfo {volume} {128}},\ \bibinfo {pages}
  {1} (\bibinfo {year} {2013})}\BibitemShut {NoStop}%
\bibitem [{\citenamefont {{Intergovernmental Panel on Climate Change
  (IPCC)}}(2014)}]{ipcc}%
  \BibitemOpen
  \bibfield  {author} {\bibinfo {author} {\bibnamefont {{Intergovernmental
  Panel on Climate Change (IPCC)}}},\ }\href@noop {} {\bibinfo {title} {Fifth
  assessment report}} (\bibinfo {year} {2014})\BibitemShut {NoStop}%
\bibitem [{\citenamefont {Abbrescia}\ \emph
  {et~al.}(2016{\natexlab{a}})\citenamefont {Abbrescia}, \citenamefont
  {Benussi}, \citenamefont {Piccolo}, \citenamefont {Bianco}, \citenamefont
  {Ferrini}, \citenamefont {Muhammad}, \citenamefont {Passamonti},
  \citenamefont {Pierluigi}, \citenamefont {Primavera}, \citenamefont {Russo}
  \emph {et~al.}}]{abbrescia2016eco}%
  \BibitemOpen
  \bibfield  {author} {\bibinfo {author} {\bibfnamefont {M.}~\bibnamefont
  {Abbrescia}}, \bibinfo {author} {\bibfnamefont {L.}~\bibnamefont {Benussi}},
  \bibinfo {author} {\bibfnamefont {D.}~\bibnamefont {Piccolo}}, \bibinfo
  {author} {\bibfnamefont {S.}~\bibnamefont {Bianco}}, \bibinfo {author}
  {\bibfnamefont {M.}~\bibnamefont {Ferrini}}, \bibinfo {author} {\bibfnamefont
  {S.}~\bibnamefont {Muhammad}}, \bibinfo {author} {\bibfnamefont
  {L.}~\bibnamefont {Passamonti}}, \bibinfo {author} {\bibfnamefont
  {D.}~\bibnamefont {Pierluigi}}, \bibinfo {author} {\bibfnamefont
  {F.}~\bibnamefont {Primavera}}, \bibinfo {author} {\bibfnamefont
  {A.}~\bibnamefont {Russo}}, \emph {et~al.},\ }\bibfield  {title} {\bibinfo
  {title} {Eco-friendly gas mixtures for {R}esistive {P}late {C}hambers based
  on tetrafluoropropene and helium},\ }\href@noop {} {\bibfield  {journal}
  {\bibinfo  {journal} {Journal of Instrumentation}\ }\textbf {\bibinfo
  {volume} {11}}\bibinfo  {number} { (08)},\ \bibinfo {pages}
  {P08019}}\BibitemShut {NoStop}%
\bibitem [{\citenamefont {Guida}\ \emph {et~al.}(2016)\citenamefont {Guida},
  \citenamefont {Capeans},\ and\ \citenamefont
  {Mandelli}}]{guida2016characterization}%
  \BibitemOpen
\bibfield  {number} {  }\bibfield  {author} {\bibinfo {author} {\bibfnamefont
  {R.}~\bibnamefont {Guida}}, \bibinfo {author} {\bibfnamefont
  {M.}~\bibnamefont {Capeans}},\ and\ \bibinfo {author} {\bibfnamefont
  {B.}~\bibnamefont {Mandelli}},\ }\bibfield  {title} {\bibinfo {title}
  {Characterization of {RPC} operation with new environmental friendly mixtures
  for {LHC} application and beyond},\ }\href@noop {} {\bibfield  {journal}
  {\bibinfo  {journal} {Journal of Instrumentation}\ }\textbf {\bibinfo
  {volume} {11}}\bibinfo  {number} { (07)},\ \bibinfo {pages}
  {C07016}}\BibitemShut {NoStop}%
\bibitem [{\citenamefont {Liberti}\ \emph {et~al.}(2016)\citenamefont
  {Liberti}, \citenamefont {Aielli}, \citenamefont {Camarri}, \citenamefont
  {Cardarelli}, \citenamefont {Di~Ciaccio}, \citenamefont {Di~Stante},
  \citenamefont {Pastori},\ and\ \citenamefont
  {Santonico}}]{liberti2016further}%
  \BibitemOpen
\bibfield  {number} {  }\bibfield  {author} {\bibinfo {author} {\bibfnamefont
  {B.}~\bibnamefont {Liberti}}, \bibinfo {author} {\bibfnamefont
  {G.}~\bibnamefont {Aielli}}, \bibinfo {author} {\bibfnamefont
  {P.}~\bibnamefont {Camarri}}, \bibinfo {author} {\bibfnamefont
  {R.}~\bibnamefont {Cardarelli}}, \bibinfo {author} {\bibfnamefont
  {A.}~\bibnamefont {Di~Ciaccio}}, \bibinfo {author} {\bibfnamefont
  {L.}~\bibnamefont {Di~Stante}}, \bibinfo {author} {\bibfnamefont
  {E.}~\bibnamefont {Pastori}},\ and\ \bibinfo {author} {\bibfnamefont
  {R.}~\bibnamefont {Santonico}},\ }\bibfield  {title} {\bibinfo {title}
  {Further gas mixtures with low environment impact},\ }\href@noop {}
  {\bibfield  {journal} {\bibinfo  {journal} {Journal of Instrumentation}\
  }\textbf {\bibinfo {volume} {11}}\bibinfo  {number} { (09)},\ \bibinfo
  {pages} {C09012}}\BibitemShut {NoStop}%
\bibitem [{\citenamefont {Bianchi}\ \emph {et~al.}(2019)\citenamefont
  {Bianchi}, \citenamefont {Delsanto}, \citenamefont {Dupieux}, \citenamefont
  {Ferretti}, \citenamefont {Gagliardi}, \citenamefont {Joly}, \citenamefont
  {Manen}, \citenamefont {Marchisone}, \citenamefont {Micheletti},
  \citenamefont {Rosano} \emph {et~al.}}]{bianchi2019characterization}%
  \BibitemOpen
\bibfield  {number} {  }\bibfield  {author} {\bibinfo {author} {\bibfnamefont
  {A.}~\bibnamefont {Bianchi}}, \bibinfo {author} {\bibfnamefont
  {S.}~\bibnamefont {Delsanto}}, \bibinfo {author} {\bibfnamefont
  {P.}~\bibnamefont {Dupieux}}, \bibinfo {author} {\bibfnamefont
  {A.}~\bibnamefont {Ferretti}}, \bibinfo {author} {\bibfnamefont
  {M.}~\bibnamefont {Gagliardi}}, \bibinfo {author} {\bibfnamefont
  {B.}~\bibnamefont {Joly}}, \bibinfo {author} {\bibfnamefont {S.}~\bibnamefont
  {Manen}}, \bibinfo {author} {\bibfnamefont {M.}~\bibnamefont {Marchisone}},
  \bibinfo {author} {\bibfnamefont {L.}~\bibnamefont {Micheletti}}, \bibinfo
  {author} {\bibfnamefont {A.}~\bibnamefont {Rosano}}, \emph {et~al.},\
  }\bibfield  {title} {\bibinfo {title} {Characterization of
  tetrafluoropropene-based gas mixtures for the {R}esistive {P}late {C}hambers
  of the {ALICE} muon spectrometer},\ }\href@noop {} {\bibfield  {journal}
  {\bibinfo  {journal} {Journal of Instrumentation}\ }\textbf {\bibinfo
  {volume} {14}}\bibinfo  {number} { (11)},\ \bibinfo {pages}
  {P11014}}\BibitemShut {NoStop}%
\bibitem [{\citenamefont {Proto}\ \emph {et~al.}(2022)\citenamefont {Proto},
  \citenamefont {Liberti}, \citenamefont {Santonico}, \citenamefont {Aielli},
  \citenamefont {Camarri}, \citenamefont {Cardarelli}, \citenamefont
  {Di~Ciaccio}, \citenamefont {Di~Stante}, \citenamefont {Paoloni},
  \citenamefont {Pastori} \emph {et~al.}}]{proto2022new}%
  \BibitemOpen
\bibfield  {number} {  }\bibfield  {author} {\bibinfo {author} {\bibfnamefont
  {G.}~\bibnamefont {Proto}}, \bibinfo {author} {\bibfnamefont
  {B.}~\bibnamefont {Liberti}}, \bibinfo {author} {\bibfnamefont
  {R.}~\bibnamefont {Santonico}}, \bibinfo {author} {\bibfnamefont
  {G.}~\bibnamefont {Aielli}}, \bibinfo {author} {\bibfnamefont
  {P.}~\bibnamefont {Camarri}}, \bibinfo {author} {\bibfnamefont
  {R.}~\bibnamefont {Cardarelli}}, \bibinfo {author} {\bibfnamefont
  {A.}~\bibnamefont {Di~Ciaccio}}, \bibinfo {author} {\bibfnamefont
  {L.}~\bibnamefont {Di~Stante}}, \bibinfo {author} {\bibfnamefont
  {A.}~\bibnamefont {Paoloni}}, \bibinfo {author} {\bibfnamefont
  {E.}~\bibnamefont {Pastori}}, \emph {et~al.},\ }\bibfield  {title} {\bibinfo
  {title} {On a new environment-friendly gas mixture for {R}esistive {P}late
  {C}hambers},\ }\href@noop {} {\bibfield  {journal} {\bibinfo  {journal}
  {Journal of Instrumentation}\ }\textbf {\bibinfo {volume} {17}}\bibinfo
  {number} { (05)},\ \bibinfo {pages} {P05005}}\BibitemShut {NoStop}%
\bibitem [{\citenamefont {Abbrescia}\ \emph
  {et~al.}(2016{\natexlab{b}})\citenamefont {Abbrescia}, \citenamefont
  {Van~Auwegem}, \citenamefont {Benussi}, \citenamefont {Bianco}, \citenamefont
  {Cauwenbergh}, \citenamefont {Ferrini}, \citenamefont {Muhammad},
  \citenamefont {Passamonti}, \citenamefont {Pierluigi}, \citenamefont
  {Piccolo} \emph {et~al.}}]{abbrescia2016preliminary}%
  \BibitemOpen
\bibfield  {number} {  }\bibfield  {author} {\bibinfo {author} {\bibfnamefont
  {M.}~\bibnamefont {Abbrescia}}, \bibinfo {author} {\bibfnamefont
  {P.}~\bibnamefont {Van~Auwegem}}, \bibinfo {author} {\bibfnamefont
  {L.}~\bibnamefont {Benussi}}, \bibinfo {author} {\bibfnamefont
  {S.}~\bibnamefont {Bianco}}, \bibinfo {author} {\bibfnamefont
  {S.}~\bibnamefont {Cauwenbergh}}, \bibinfo {author} {\bibfnamefont
  {M.}~\bibnamefont {Ferrini}}, \bibinfo {author} {\bibfnamefont
  {S.}~\bibnamefont {Muhammad}}, \bibinfo {author} {\bibfnamefont
  {L.}~\bibnamefont {Passamonti}}, \bibinfo {author} {\bibfnamefont
  {D.}~\bibnamefont {Pierluigi}}, \bibinfo {author} {\bibfnamefont
  {D.}~\bibnamefont {Piccolo}}, \emph {et~al.},\ }\bibfield  {title} {\bibinfo
  {title} {Preliminary results of {R}esistive {P}late {C}hambers operated with
  eco-friendly gas mixtures for application in the {CMS} experiment},\
  }\href@noop {} {\bibfield  {journal} {\bibinfo  {journal} {Journal of
  Instrumentation}\ }\textbf {\bibinfo {volume} {11}}\bibinfo  {number} {
  (09)},\ \bibinfo {pages} {C09018}}\BibitemShut {NoStop}%
\bibitem [{\citenamefont {Rigoletti}\ \emph {et~al.}(2020)\citenamefont
  {Rigoletti}, \citenamefont {Aielli}, \citenamefont {Alberghi}, \citenamefont
  {Benussi}, \citenamefont {Bianchi}, \citenamefont {Bianco}, \citenamefont
  {Di~Stante}, \citenamefont {Boscherini}, \citenamefont {Bruni}, \citenamefont
  {Camarri} \emph {et~al.}}]{rigoletti2020studies}%
  \BibitemOpen
\bibfield  {number} {  }\bibfield  {author} {\bibinfo {author} {\bibfnamefont
  {G.}~\bibnamefont {Rigoletti}}, \bibinfo {author} {\bibfnamefont
  {G.}~\bibnamefont {Aielli}}, \bibinfo {author} {\bibfnamefont
  {G.}~\bibnamefont {Alberghi}}, \bibinfo {author} {\bibfnamefont
  {L.}~\bibnamefont {Benussi}}, \bibinfo {author} {\bibfnamefont
  {A.}~\bibnamefont {Bianchi}}, \bibinfo {author} {\bibfnamefont
  {S.}~\bibnamefont {Bianco}}, \bibinfo {author} {\bibfnamefont
  {L.}~\bibnamefont {Di~Stante}}, \bibinfo {author} {\bibfnamefont
  {D.}~\bibnamefont {Boscherini}}, \bibinfo {author} {\bibfnamefont
  {A.}~\bibnamefont {Bruni}}, \bibinfo {author} {\bibfnamefont
  {P.}~\bibnamefont {Camarri}}, \emph {et~al.},\ }\bibfield  {title} {\bibinfo
  {title} {Studies of {RPC} detector operation with eco-friendly gas mixtures
  under irradiation at the {CERN} {G}amma {I}rradiation {F}acility},\
  }\href@noop {} {\bibfield  {journal} {\bibinfo  {journal} {POS Proceedings of
  Science}\ }\textbf {\bibinfo {volume} {364}},\ \bibinfo {pages} {164}
  (\bibinfo {year} {2020})}\BibitemShut {NoStop}%
\bibitem [{\citenamefont {Chachereau}\ \emph {et~al.}(2016)\citenamefont
  {Chachereau}, \citenamefont {Rabie},\ and\ \citenamefont
  {Franck}}]{chachereau2016electron}%
  \BibitemOpen
  \bibfield  {author} {\bibinfo {author} {\bibfnamefont {A.}~\bibnamefont
  {Chachereau}}, \bibinfo {author} {\bibfnamefont {M.}~\bibnamefont {Rabie}},\
  and\ \bibinfo {author} {\bibfnamefont {C.~M.}\ \bibnamefont {Franck}},\
  }\bibfield  {title} {\bibinfo {title} {Electron swarm parameters of the
  hydrofluoroolefine hfo1234ze},\ }\href@noop {} {\bibfield  {journal}
  {\bibinfo  {journal} {Plasma Sources Science and Technology}\ }\textbf
  {\bibinfo {volume} {25}},\ \bibinfo {pages} {045005} (\bibinfo {year}
  {2016})}\BibitemShut {NoStop}%
\bibitem [{\citenamefont {Bianchi}\ \emph {et~al.}(2020)\citenamefont
  {Bianchi}, \citenamefont {Delsanto}, \citenamefont {Dupieux}, \citenamefont
  {Ferretti}, \citenamefont {Gagliardi}, \citenamefont {Joly}, \citenamefont
  {Manen}, \citenamefont {Marchisone}, \citenamefont {Micheletti},
  \citenamefont {Rosano} \emph {et~al.}}]{bianchi2020studies}%
  \BibitemOpen
  \bibfield  {author} {\bibinfo {author} {\bibfnamefont {A.}~\bibnamefont
  {Bianchi}}, \bibinfo {author} {\bibfnamefont {S.}~\bibnamefont {Delsanto}},
  \bibinfo {author} {\bibfnamefont {P.}~\bibnamefont {Dupieux}}, \bibinfo
  {author} {\bibfnamefont {A.}~\bibnamefont {Ferretti}}, \bibinfo {author}
  {\bibfnamefont {M.}~\bibnamefont {Gagliardi}}, \bibinfo {author}
  {\bibfnamefont {B.}~\bibnamefont {Joly}}, \bibinfo {author} {\bibfnamefont
  {S.}~\bibnamefont {Manen}}, \bibinfo {author} {\bibfnamefont
  {M.}~\bibnamefont {Marchisone}}, \bibinfo {author} {\bibfnamefont
  {L.}~\bibnamefont {Micheletti}}, \bibinfo {author} {\bibfnamefont
  {A.}~\bibnamefont {Rosano}}, \emph {et~al.},\ }\bibfield  {title} {\bibinfo
  {title} {Studies on tetrafluoropropene-based gas mixtures with low
  environmental impact for {R}esistive {P}late {C}hambers},\ }\href@noop {}
  {\bibfield  {journal} {\bibinfo  {journal} {Journal of Instrumentation}\
  }\textbf {\bibinfo {volume} {15}}\bibinfo  {number} { (04)},\ \bibinfo
  {pages} {C04039}}\BibitemShut {NoStop}%
\bibitem [{\citenamefont {Biagi}(1999)}]{biagi1999monte}%
  \BibitemOpen
\bibfield  {number} {  }\bibfield  {author} {\bibinfo {author} {\bibfnamefont
  {S.}~\bibnamefont {Biagi}},\ }\bibfield  {title} {\bibinfo {title} {{M}onte
  {C}arlo simulation of electron drift and diffusion in counting gases under
  the influence of electric and magnetic fields},\ }\href@noop {} {\bibfield
  {journal} {\bibinfo  {journal} {Nuclear Instruments and Methods in Physics
  Research Section A}\ }\textbf {\bibinfo {volume} {421}},\ \bibinfo {pages}
  {234} (\bibinfo {year} {1999})}\BibitemShut {NoStop}%
\bibitem [{\citenamefont {Bianchi}\ \emph {et~al.}(2021)\citenamefont
  {Bianchi}, \citenamefont {Ferretti}, \citenamefont {Gagliardi},\ and\
  \citenamefont {Vercellin}}]{bianchi2021electron}%
  \BibitemOpen
  \bibfield  {author} {\bibinfo {author} {\bibfnamefont {A.}~\bibnamefont
  {Bianchi}}, \bibinfo {author} {\bibfnamefont {A.}~\bibnamefont {Ferretti}},
  \bibinfo {author} {\bibfnamefont {M.}~\bibnamefont {Gagliardi}},\ and\
  \bibinfo {author} {\bibfnamefont {E.}~\bibnamefont {Vercellin}},\ }\bibfield
  {title} {\bibinfo {title} {Electron collision cross sections in
  tetrafluoropropene {HFO}1234ze},\ }\href@noop {} {\bibfield  {journal}
  {\bibinfo  {journal} {arXiv preprint arXiv:2103.08643}\ } (\bibinfo {year}
  {2021})}\BibitemShut {NoStop}%
\bibitem [{\citenamefont {Rabie}\ and\ \citenamefont
  {Franck}(2016)}]{rabie2016methes}%
  \BibitemOpen
  \bibfield  {author} {\bibinfo {author} {\bibfnamefont {M.}~\bibnamefont
  {Rabie}}\ and\ \bibinfo {author} {\bibfnamefont {C.~M.}\ \bibnamefont
  {Franck}},\ }\bibfield  {title} {\bibinfo {title} {{METHES}: {A} {M}onte
  {C}arlo collision code for the simulation of electron transport in low
  temperature plasmas},\ }\href@noop {} {\bibfield  {journal} {\bibinfo
  {journal} {Computer Physics Communications}\ }\textbf {\bibinfo {volume}
  {203}},\ \bibinfo {pages} {268} (\bibinfo {year} {2016})}\BibitemShut
  {NoStop}%
\bibitem [{\citenamefont {Pancheshnyi}\ \emph {et~al.}(2012)\citenamefont
  {Pancheshnyi}, \citenamefont {Biagi}, \citenamefont {Bordage}, \citenamefont
  {Hagelaar}, \citenamefont {Morgan}, \citenamefont {Phelps},\ and\
  \citenamefont {Pitchford}}]{lxcat_project}%
  \BibitemOpen
  \bibfield  {author} {\bibinfo {author} {\bibfnamefont {S.}~\bibnamefont
  {Pancheshnyi}}, \bibinfo {author} {\bibfnamefont {S.}~\bibnamefont {Biagi}},
  \bibinfo {author} {\bibfnamefont {M.}~\bibnamefont {Bordage}}, \bibinfo
  {author} {\bibfnamefont {G.}~\bibnamefont {Hagelaar}}, \bibinfo {author}
  {\bibfnamefont {W.}~\bibnamefont {Morgan}}, \bibinfo {author} {\bibfnamefont
  {A.}~\bibnamefont {Phelps}},\ and\ \bibinfo {author} {\bibfnamefont
  {L.}~\bibnamefont {Pitchford}},\ }\bibfield  {title} {\bibinfo {title} {The
  {LXC}at project: Electron scattering cross sections and swarm parameters for
  low temperature plasma modeling},\ }\href@noop {} {\bibfield  {journal}
  {\bibinfo  {journal} {Chemical Physics}\ }\textbf {\bibinfo {volume} {398}},\
  \bibinfo {pages} {148} (\bibinfo {year} {2012})}\BibitemShut {NoStop}%
\bibitem [{\citenamefont {Dagum}\ and\ \citenamefont {Menon}(1998)}]{openmp}%
  \BibitemOpen
  \bibfield  {author} {\bibinfo {author} {\bibfnamefont {L.}~\bibnamefont
  {Dagum}}\ and\ \bibinfo {author} {\bibfnamefont {R.}~\bibnamefont {Menon}},\
  }\bibfield  {title} {\bibinfo {title} {Open{MP}: an industry standard {API}
  for shared-memory programming},\ }\href@noop {} {\bibfield  {journal}
  {\bibinfo  {journal} {IEEE computational science and engineering}\ }\textbf
  {\bibinfo {volume} {5}},\ \bibinfo {pages} {46} (\bibinfo {year}
  {1998})}\BibitemShut {NoStop}%
\bibitem [{\citenamefont {Antcheva}\ \emph {et~al.}(2011)\citenamefont
  {Antcheva}, \citenamefont {Ballintijn}, \citenamefont {Bellenot},
  \citenamefont {Biskup}, \citenamefont {Brun}, \citenamefont {Buncic},
  \citenamefont {Canal}, \citenamefont {Casadei}, \citenamefont {Couet},
  \citenamefont {Fine} \emph {et~al.}}]{root_documentation}%
  \BibitemOpen
  \bibfield  {author} {\bibinfo {author} {\bibfnamefont {I.}~\bibnamefont
  {Antcheva}}, \bibinfo {author} {\bibfnamefont {M.}~\bibnamefont
  {Ballintijn}}, \bibinfo {author} {\bibfnamefont {B.}~\bibnamefont
  {Bellenot}}, \bibinfo {author} {\bibfnamefont {M.}~\bibnamefont {Biskup}},
  \bibinfo {author} {\bibfnamefont {R.}~\bibnamefont {Brun}}, \bibinfo {author}
  {\bibfnamefont {N.}~\bibnamefont {Buncic}}, \bibinfo {author} {\bibfnamefont
  {P.}~\bibnamefont {Canal}}, \bibinfo {author} {\bibfnamefont
  {D.}~\bibnamefont {Casadei}}, \bibinfo {author} {\bibfnamefont
  {O.}~\bibnamefont {Couet}}, \bibinfo {author} {\bibfnamefont
  {V.}~\bibnamefont {Fine}}, \emph {et~al.},\ }\bibfield  {title} {\bibinfo
  {title} {{ROOT}: {A} {C}++ framework for petabyte data storage, statistical
  analysis and visualization},\ }\href@noop {} {\bibfield  {journal} {\bibinfo
  {journal} {Computer Physics Communications}\ }\textbf {\bibinfo {volume}
  {182}},\ \bibinfo {pages} {1384} (\bibinfo {year} {2011})}\BibitemShut
  {NoStop}%
\bibitem [{\citenamefont {Abbrescia}\ \emph {et~al.}(2018)\citenamefont
  {Abbrescia}, \citenamefont {Fonte},\ and\ \citenamefont
  {Peskov}}]{abbrescia_book}%
  \BibitemOpen
  \bibfield  {author} {\bibinfo {author} {\bibfnamefont {M.}~\bibnamefont
  {Abbrescia}}, \bibinfo {author} {\bibfnamefont {P.}~\bibnamefont {Fonte}},\
  and\ \bibinfo {author} {\bibfnamefont {V.}~\bibnamefont {Peskov}},\
  }\href@noop {} {\emph {\bibinfo {title} {{{R}esistive {G}aseous {D}etectors:
  designs, performance, and perspectives}}}}\ (\bibinfo  {publisher}
  {Wiley-VCH},\ \bibinfo {address} {Weinheim},\ \bibinfo {year}
  {2018})\BibitemShut {NoStop}%
\bibitem [{\citenamefont {Zeballos}\ \emph
  {et~al.}(1996{\natexlab{b}})\citenamefont {Zeballos}, \citenamefont {Crotty},
  \citenamefont {Hatzifotiadou}, \citenamefont {Valverde}, \citenamefont
  {Neupane}, \citenamefont {Williams},\ and\ \citenamefont {Zichichi}}]{mrpc}%
  \BibitemOpen
  \bibfield  {author} {\bibinfo {author} {\bibfnamefont {E.~C.}\ \bibnamefont
  {Zeballos}}, \bibinfo {author} {\bibfnamefont {I.}~\bibnamefont {Crotty}},
  \bibinfo {author} {\bibfnamefont {D.}~\bibnamefont {Hatzifotiadou}}, \bibinfo
  {author} {\bibfnamefont {J.~L.}\ \bibnamefont {Valverde}}, \bibinfo {author}
  {\bibfnamefont {S.}~\bibnamefont {Neupane}}, \bibinfo {author} {\bibfnamefont
  {M.}~\bibnamefont {Williams}},\ and\ \bibinfo {author} {\bibfnamefont
  {A.}~\bibnamefont {Zichichi}},\ }\bibfield  {title} {\bibinfo {title} {A new
  type of {R}esistive {P}late {C}hamber: the multigap {RPC}},\ }\href@noop {}
  {\bibfield  {journal} {\bibinfo  {journal} {Nuclear Instruments and Methods
  in Physics Research Section A}\ }\textbf {\bibinfo {volume} {374}},\ \bibinfo
  {pages} {132} (\bibinfo {year} {1996}{\natexlab{b}})}\BibitemShut {NoStop}%
\bibitem [{\citenamefont {Lippmann}\ and\ \citenamefont
  {Riegler}(2004)}]{lippmann2004space}%
  \BibitemOpen
  \bibfield  {author} {\bibinfo {author} {\bibfnamefont {C.}~\bibnamefont
  {Lippmann}}\ and\ \bibinfo {author} {\bibfnamefont {W.}~\bibnamefont
  {Riegler}},\ }\bibfield  {title} {\bibinfo {title} {Space charge effects in
  {R}esistive {P}late {C}hambers},\ }\href@noop {} {\bibfield  {journal}
  {\bibinfo  {journal} {Nuclear Instruments and Methods in Physics Research
  Section A: Accelerators, Spectrometers, Detectors and Associated Equipment}\
  }\textbf {\bibinfo {volume} {517}},\ \bibinfo {pages} {54} (\bibinfo {year}
  {2004})}\BibitemShut {NoStop}%
\bibitem [{\citenamefont {Skullerud}(1968)}]{null1}%
  \BibitemOpen
  \bibfield  {author} {\bibinfo {author} {\bibfnamefont {H.}~\bibnamefont
  {Skullerud}},\ }\bibfield  {title} {\bibinfo {title} {The stochastic computer
  simulation of ion motion in a gas subjected to a constant electric field},\
  }\href@noop {} {\bibfield  {journal} {\bibinfo  {journal} {Journal of Physics
  D: Applied Physics}\ }\textbf {\bibinfo {volume} {1}},\ \bibinfo {pages}
  {1567} (\bibinfo {year} {1968})}\BibitemShut {NoStop}%
\bibitem [{\citenamefont {Koura}(1986)}]{null2}%
  \BibitemOpen
  \bibfield  {author} {\bibinfo {author} {\bibfnamefont {K.}~\bibnamefont
  {Koura}},\ }\bibfield  {title} {\bibinfo {title} {Null-collision technique in
  the direct-simulation {M}onte {C}arlo method},\ }\href@noop {} {\bibfield
  {journal} {\bibinfo  {journal} {The Physics of fluids}\ }\textbf {\bibinfo
  {volume} {29}},\ \bibinfo {pages} {3509} (\bibinfo {year}
  {1986})}\BibitemShut {NoStop}%
\bibitem [{\citenamefont {Landau}\ and\ \citenamefont
  {Lifshitz}(2013)}]{landau}%
  \BibitemOpen
  \bibfield  {author} {\bibinfo {author} {\bibfnamefont {L.~D.}\ \bibnamefont
  {Landau}}\ and\ \bibinfo {author} {\bibfnamefont {E.~M.}\ \bibnamefont
  {Lifshitz}},\ }\href@noop {} {\emph {\bibinfo {title} {Mechanics and
  electrodynamics}}}\ (\bibinfo  {publisher} {Elsevier},\ \bibinfo {year}
  {2013})\BibitemShut {NoStop}%
\bibitem [{\citenamefont {{\v{S}}a{\v{s}}i{\'c}}\ \emph
  {et~al.}(2013)\citenamefont {{\v{S}}a{\v{s}}i{\'c}}, \citenamefont
  {Dupljanin}, \citenamefont {de~Urquijo},\ and\ \citenamefont
  {Petrovi{\'c}}}]{vsavsicswarm}%
  \BibitemOpen
  \bibfield  {author} {\bibinfo {author} {\bibfnamefont {O.}~\bibnamefont
  {{\v{S}}a{\v{s}}i{\'c}}}, \bibinfo {author} {\bibfnamefont {S.}~\bibnamefont
  {Dupljanin}}, \bibinfo {author} {\bibfnamefont {J.}~\bibnamefont
  {de~Urquijo}},\ and\ \bibinfo {author} {\bibfnamefont {Z.~L.}\ \bibnamefont
  {Petrovi{\'c}}},\ }\bibfield  {title} {\bibinfo {title} {Scattering cross
  sections for electrons in {C2H2F4} and its mixtures with {A}r from measured
  transport coefficients},\ }\href@noop {} {\bibfield  {journal} {\bibinfo
  {journal} {Journal of Physics D: Applied Physics}\ }\textbf {\bibinfo
  {volume} {46}},\ \bibinfo {pages} {325201} (\bibinfo {year}
  {2013})}\BibitemShut {NoStop}%
\bibitem [{\citenamefont {{The European Parliament and the
  Council}}(2014)}]{europeanparliament}%
  \BibitemOpen
  \bibfield  {author} {\bibinfo {author} {\bibnamefont {{The European
  Parliament and the Council}}},\ }\bibfield  {title} {\bibinfo {title}
  {Regulation ({EU}) no 517/2014 on fluorinated greenhouse gases},\ }\href@noop
  {} {\bibfield  {journal} {\bibinfo  {journal} {Official Journal of the
  European Union}\ }\textbf {\bibinfo {volume} {L 150}},\ \bibinfo {pages}
  {195} (\bibinfo {year} {2014})}\BibitemShut {NoStop}%
\bibitem [{lxc()}]{lxcat_database}%
  \BibitemOpen
  \href@noop {} {}\bibinfo {note} {Biagi's database, www.lxcat.net, retrieved
  in January, 2020}\BibitemShut {NoStop}%
\end{thebibliography}%

\end{document}